\documentclass[aps,prev,12pt,preprintnumbers,floatfix,nofootinbib]{revtex4-1}
\usepackage{graphicx}
\usepackage{bm}
\usepackage{times}
\usepackage{amsmath}
\usepackage{amsfonts}
\usepackage{amssymb}
\usepackage{slashed}
\usepackage{color}
\usepackage{epstopdf}
\usepackage{subfigure}
\usepackage{xfrac}

\def\br{\begin{eqnarray}}
\def\er{\end{eqnarray}}
\def\be{\begin{equation}}
\def\ee{\end{equation}}

\def\({\left(}
\def\){\right)}

\newcommand \beq { \begin{eqnarray} }
\newcommand \eeq { \end{eqnarray} }
\newcommand \beqq { \begin{equation} }
\newcommand \eeqq { \end{equation} }

\begin{document}

%\title{New Physics Contributions to the Muon Anomalous Magnetic Moment}
\title{New Physics Contributions to the Muon Anomalous Magnetic Moment: A Numerical Code}
\vspace{2cm}
\author{Farinaldo S. Queiroz}
\author{William Shepherd}
\vspace{2cm}
\affiliation{Department of Physics and Santa Cruz Institute for Particle Physics
University of California, Santa Cruz, CA 95064, USA}

\vspace{2cm}
\begin{abstract}
We consider the contributions of individual new particles to the anomalous magnetic moment of the muon, utilizing the generic framework of simplified models. We also present analytic results for all possible one-loop contributions, allowing easy application of these results for more complete models which predict more than one particle capable of correcting the muon magnetic moment. Additionally, we provide a Mathematica code to allow the reader straightforwardly compute any 1-loop contribution. Furthermore, we derive bounds on each new particle considered, assuming either the absence of other significant contributions to $a_\mu$ or that the anomaly has been resolved by some other mechanism. The simplified models we consider are constructed without the requirement of $SU(2)_L$ invariance, but appropriate chiral coupling choices are also considered. In summary, we found the following particles capable of explaining the current discrepancy, assuming unit couplings: $2$~TeV ($0.3$~TeV) neutral scalar with pure scalar (chiral) couplings, $4$~TeV doubly charged scalar with pure pseudoscalar coupling, $0.3-1$~TeV neutral vector boson depending on what couplings are used (vector, axial, or mixed), $0.5-1$~TeV singly-charged vector boson depending on which couplings are chosen, and $3$~TeV doubly-charged vector-coupled bosons.
We also derive the following $1\sigma$ lower bounds on new particle masses assuming unit couplings and that the experimental anomaly has been otherwise resolved: a doubly charged pseudo-scalar must be heavier than $7$~TeV, a neutral scalar than $3$~TeV, a vector-coupled new neutral boson $600$~GeV, an axial-coupled neutral boson $1.5$~TeV, a singly-charged vector-coupled $W^\prime$ $1$~TeV, a doubly-charged vector-coupled boson $5$~TeV, scalar leptoquarks $10$~TeV, and vector leptoquarks $10$~TeV. We emphasize that the quoted numbers apply within simplified models, but the reader can easily use our Mathemata code to calculate the contribution of their own model of new physics.

%\begin{tabular}{|c|c|}
%\hline
%Pure pseudo-doubly charged scalar & $7$~TeV \\
%Neutral scalar & $4$~TeV \\
%Pure neutral vector boson & $600$~GeV \\
%Pure neutral vector boson & $1$~TeV\\
%Singly charged Vector & $1$~TeV\\
%Doubly charged Vector & $5$~TeV\\
%Scalar leptoquarks & $10$~TeV\\
%Pure vector leptoquarks & $10$~TeV\\
%\hline
%\end{tabular}

\end{abstract}
   
\maketitle

\section{Introduction}

The muon anomalous magnetic moment, measured to amazing precision, is currently one of the most compelling inconsistencies between data and the Standard Model (SM) predictions in all of particle physics. With current data the discrepancy is at the $3.6\sigma$ level \cite{PDG}, and efforts are underway on both the experimental \cite{fermilabproposal} and theoretical fronts 
\cite{g2improvements1,g2improvements2,g2improvements3,g2improvements4,g2improvements5,g2improvements6,g2improvements7} to improve the precision of both the measured value and the SM prediction. This deviation from the expectations of the SM has been used as motivation for many models of new physics \cite{g2models1,g2models2,g2models3,g2models4,g2models5,g2models6,g2models7,Agrawal:2014ufa}, and has been used to constrain or motivate parameter values for many others, notably supersymmetric models \cite{g2models8,g2models9,g2models10,g2models11,g2models12}.

The literature on this topic has been somewhat scattershot, with many authors focusing solely on their preferred model to explain (or be constrained by) this data. Here, we aim to provide a more complete and reusable lexicon for the use of the community. We proceed using the generic framework of simplified models \cite{g2generic1,g2generic2,g2generic3,g2generic4,g2generic5,g2generic6}, assuming the existence of a minimal number of particles that could have the phenomenological impact we are interested in. We do not enforce $SU(2)_L$ invariance in the simplified models we construct, as the CP basis is more relevant to the calculation of low energy phenomena and gives linearly independent 'basis element' contributions in a way that chiral couplings do not. There are, naturally, a very large number of such simplified models which contribute to the muon anomalous magnetic moment, but we consider a basis sufficient to parametrize the contribution of any new particle with spin $s\le1$.

We provide bounds on the couplings and masses of each of our 'basis element' models by requiring sufficient consistency with the experimental value of $a_\mu$. Naturally, a model which is some combination of our basis elements will be subject to slightly different bounds, and as those models are not intrinsically less interesting than our basis models alone, we provide the needed analytic formulae to allow the derivation of the appropriate bounds on any model which can contribute at one-loop order to the magnetic moment.  We note that this will provide the ability to consider not only different coupling structures but also models in which multiple particles can individually contribute. While analytic results have been provided previously, there are some disagreements in the literature. Also, we will present for the first time results which include all effects due to the finite muon mass. These results have also been provided in the form of a Mathematica notebook, which calculates the exact contribution to $a_\mu$ from any one-loop new physics diagram.

This paper is organized as follows: in the next section we state the rules we use to select a basis of simplified models that contribute to $\Delta a_\mu$ and present the models themselves. In section \ref{sec:calc} we calculate the contribution of each model to the muon magnetic moment and present analytic formulae for those contributions, as well as bounds on each simplified model based on our Public Mathematica code. Finally, in section \ref{sec:conc} we conclude. Analytic formulae derived without taking limits relating to the masses of new particles are presented in appendix \ref{appendix}.

\section{Simplified Models of $\Delta a_\mu$}
\label{sec:mod}

We construct our main set of simplified models by requiring that they contribute at one-loop order to the muon anomalous magnetic moment, and that they do so with only one new particle. While contributions which involve two new particles are certainly possible and interesting, their contributions can be calculated from the same master integrals we provide as these. We further restrict ourselves to consider only particles of spin $s\le1$, in the interest of concreteness. Finally, we neglect any interactions which violate lepton flavor when the new particle is not itself a new lepton. Any flavor-violating interactions have effects which are similar to the flavor-conserving interactions we consider here \cite{g2flavor1,g2flavor2}.

We can further categorize the contributions in terms of the charges of the new particle. Subject to the above constraints, a colorless new particle can only contribute to $a_\mu$ if it is neutral or has unit or double-unit charge. In this way, we have identified nine different classes of possible colorless contributors to the muon anomalous magnetic moment. A colored particle can contribute as a leptoquark, but only if it is a boson. In the leptoquark case the possible charges are \sfrac{1}{3}, \sfrac{2}{3}, \sfrac{4}{3}, or \sfrac{5}{3}. There are thus eight leptoquarks we must consider.

The only remaining choices to be made are regarding the structure of the couplings of these particles to the SM fields. As we are interested in the least suppressed contributions, we will consider only renormalizable couplings. We will explore the possible couplings of each candidate class of new particles in turn. As the simplified models we consider are not expected to be complete we will not enforce various SM symmetries on the couplings. In particular, we will not require the conservation of lepton number, and we will not enforce $SU(2)_L$ invariance in the couplings, as the particles could, at least in principle, result as the mixture of different multiplets of $SU(2)_L$ in a more complete theory. A simple example would be the consideration of possible left-right mixing in the sleptons, giving a state which is not fully singlet or doublet under $SU(2)_L$. Note that it is straightfoward to return to an $SU(2)_L$ invariant coupling structure within any given simplified model by choosing a chiral combination of either scalar and pseudoscalar or vector and axial couplings; we consider these combinations as well as the basis elements alone when we interpret the contributions below.

Finally, we consider as an example two simple cases of two-particle contributions to $a_\mu$. While far from a complete exploration of all possible two-particle contributions, they will suffice to make clear how the calculations from the one-particle cases are easily generalized to allow for two-particle contributions. These two cases are also calculated in detail in the Mathematica code, allowing the reader to generalize them to their own cases of interest.

\subsection{New Colorless Scalars}

For any new scalar that is to contribute at first order to $a_\mu$ the only couplings of interest are those to the SM leptons. There are, of course, many possible such couplings. We can sort the couplings of interest by the charge of the new scalar. Considering first neutral scalars, we find that the possible interactions of interest are
\beq
{\cal L}\supset g_{s1} \phi \bar\mu\mu+ig_{p1} \phi \bar\mu\gamma^5\mu,
\label{scalar1}
\eeq
where $h$ is the new scalar of interest and $g_i$ are the couplings of each operator. Any neutral scalar can only contribute to the magnetic moment through the Feynman graph shown in Figure \ref{fig:scalar2}.

If we instead consider scalars of unit charge, we find that the relevant interactions are now
\beq
{\cal L}\supset g_{s2}\phi^+\bar\nu\mu+ ig_{p2}\phi^+\bar\nu\gamma^5\mu +g_{s3} \phi^+\bar{\nu^c}\mu+i g_{p3} \phi^+\bar{\nu^c}\gamma^5\mu+{\rm h.c.},
\label{scalar2}
\eeq
where $\psi^c$ is a charge-conjugated fermion field. These candidates can only contribute through the graph shown in Figure \ref{fig:scalar1}. The possibility of lepton number violation implied by the charge conjugation in the latter two operators, while interesting, does not affect the contribution to $a_\mu$. Therefore, we will present only one result for the charged scalars, valid for both classes of operators.

Finally, we come to consider a scalar of twice-unit charge. This candidate's interactions are
\beq
{\cal L}\supset g_{s4}\phi^{++}\bar{\mu^c}\mu+g_{p4} \phi^{++}\bar{\mu^c}\gamma^5\mu +{\rm  h.c.}.
\label{scalar3}
\eeq
Uniquely, these candidates contribute through both types of diagrams for scalar particles (Figures \ref{fig:scalar1}-\ref{fig:scalar2}). Now we will turn our attention to possible new leptons.

\subsection{New Leptons}

New leptons will contribute only through interaction terms which couple one new lepton to a muon and a SM boson. As a fourth generation of chiral fermions is forbidden by the observed couplings of the Higgs boson, particularly that to gluons, these leptons will necessarily be vector-like. Any coupling, therefore, of the type above will result only from the mixing of the new, vector-like lepton with the SM leptons. These mixings can be bounded by other flavor observables, but our focus here is on their contribution to the magnetic moment of the muon.

We will categorize the contributions of new leptons by their charge, which must again be either zero, unit, or twice-unit. For the neutral new lepton, which we will denote as $N$, the possible interactions of interest are
\beq
{\cal L}\supset g_{v5} W^+_\mu\bar N\gamma^\mu\mu+g_{a5}W^+_\mu\bar N\gamma^\mu\gamma^5\mu+{\rm h.c.}.
\label{lepton1}
\eeq
We note that these interactions have been organized in terms of vector and axial rather than left- and right-chiral couplings. This is in keeping with our previous comment that we would not be enforcing $SU(2)_L$ on our interactions, and utilizes couplings in the CP basis, which are of most relevance for the calculation of this observable, as we will discuss below. This candidate contributes through the Feynman graph shown in Figure \ref{fig:vector1}.

The next possibility is a new lepton with unit charge, which we denote as $E$. Its possible interactions are
\beq
{\cal L}\supset g_{v6} Z_\mu\bar\mu\gamma^\mu E+g_{a6} Z_\mu\bar\mu\gamma^\mu\gamma^5 E+g_{s7} h \bar\mu E+ig_{p7} h \bar\mu \gamma^5 E+{\rm h.c.}.
\label{lepton2}
\eeq
This candidate contributes through the graphs shown in Figures \ref{fig:scalar2} and \ref{fig:vector2}, depending on which interaction is considered. Once again, the possibility of lepton number violation, while independently interesting, does not affect the results for $a_\mu$.

The final new lepton candidate has charge -2, and we will denote it as $\psi$. Its interactions of interest are
\beq
{\cal L} \supset g_{v8} W^+_\mu\bar\mu\gamma^\mu\psi +g_{a8}W^+_\mu\bar\mu\gamma^\mu\gamma^5\psi+{\rm h.c.},
\label{lepton3}
\eeq
and as was the case for a doubly-charged scalar, it contributes through two diagrams as well, those shown in Figures \ref{fig:vector1} and \ref{fig:vector2}.

\subsection{New Colorless Vectors}

Much like contributions due to scalar particles, new vector particles can only contribute to the interaction through their coupling to the SM leptons. With the customary sorting by the charge of the vector boson, we begin with a new neutral vector boson, which we denote as $Z^\prime$. The relevant interactions, ignoring flavor changing couplings as discussed previously, are
\beq
{\cal L}\supset g_{v9} Z^\prime_\mu\bar\mu\gamma^\mu\mu +g_{a9} Z^\prime_\mu\bar\mu\gamma^\mu\gamma^5\mu.
\label{vector1}
\eeq
These contribute through the feynman graph shown in Figure \ref{fig:vector2}.

Moving on to charged vector bosons, which we name $W^{\prime+}$, we have
\beq
{\cal L}\supset g_{v10}W^{\prime+}_\mu\bar\nu\gamma^\mu\mu+  g_{a10}W^{\prime+}_\mu\bar\nu\gamma^\mu\gamma^5\mu+
g_{v11}W^{\prime+}_\mu\bar{\nu^c}\gamma^\mu\mu+ g_{a11}W^{\prime+}_\mu\bar{\nu^c}\gamma^\mu\gamma^5\mu+{\rm h.c.}.
\label{vector2}
\eeq
This candidate contributes only through the graph shown in Figure \ref{fig:vector1}, and once more the charge conjugation matrices are irrelevant from the point of view of the magnetic moment.

The final vector candidate is a doubly-charged one, denoted here as $U^{++}$. Its interactions of interest are
\beq
{\cal L}\supset g_{v12}U^{++}_\mu\bar{\mu^c}\gamma^\mu\mu+ g_{a12}U^{++}_\mu\bar{\mu^c}\gamma^\mu\gamma^5\mu+{\rm h.c.}.
\label{vector3}
\eeq
As has been the case for every other doubly-charged candidate, this vector boson contributes through two diagrams, those in Figures \ref{fig:vector1} and \ref{fig:vector2}.

\subsection{Scalar Leptoquarks}

Scalar leptoquarks which contribute at one loop to the anomalous moment can have charges ranging from \sfrac{1}{3} to \sfrac{5}{3}. We will discuss each in turn, distinguishing them by charge and denoting them as $\Phi^q$, with color indices suppressed. We will write our interaction terms such that the leptoquark is always a color fundamental. All scalar leptoquark candidates contribute through the diagrams shown in Figures \ref{fig:scalar1} and \ref{fig:scalar2}.

First we consider the case of charge \sfrac{1}{3} leptoquarks. The relevant interactions are
\beq
{\cal L}\supset g_{s13}\Phi^{\sfrac{-1}{3}}\bar\mu u^c+ g_{p13}\Phi^{\sfrac{-1}{3}}\bar\mu\gamma^5u^c+{\rm h.c.},
\label{leptoquarks1}
\eeq
where $u$ is an up-type quark field of (at this point) arbitrary flavor. Note that we are working with 4-component spinors here, and $u$ is not constrained to be only the $SU(2)_L$ singlet it is often used to denote in other contexts. The flavor dependence of the contribution will be explored in section \ref{sec:calc}.

If we instead postulate a charge \sfrac{2}{3} leptoquark, the interactions of interest are
\beq
{\cal L}\supset g_{s14}\Phi^{\sfrac{2}{3}}\bar d\mu+ g_{p14}\Phi^{\sfrac{2}{3}}\bar d\gamma^5\mu+{\rm h.c.},
\label{leptoquarks2}
\eeq
where $d$ is a down-type quark of indeterminate flavor, and the same caveat regarding spinor usage applies from above.

The case of charge \sfrac{4}{3} has interactions
\beq
{\cal L}\supset g_{s15}\Phi^{\sfrac{-4}{3}}\bar\mu d^c+ g_{p15}\Phi^{\sfrac{-4}{3}}\bar\mu\gamma^5d^c+{\rm h.c.},
\label{leptoquarks3}
\eeq
and the final candate with charge \sfrac{5}{3} interacts through
\beq
{\cal L}\supset g_{s16}\Phi^{\sfrac{5}{3}}\bar u\mu+ g_{p16}\Phi^{\sfrac{5}{3}}\bar u\gamma^5\mu+{\rm h.c.}.
\label{leptoquarks4}
\eeq
\subsection{Vector Leptoquarks}

Vector leptoquarks are expected to be associated with a new gauge symmetry such as a GUT, but in principle may also arise from other new physics, for instance as composite particles resulting from some new strongly-coupled interaction. They are subject to all the same constraints we required of the scalar leptoquarks, and we will adopt a similar notation for them, writing a vector leptoquark as $V^q$, with color index suppressed, always a fundamental rather than anti-fundamental.

For a vector leptoquark of charge \sfrac{1}{3}, the interactions of interest are
\beq
{\cal L}\supset g_{v17} V^{\sfrac{-1}{3}}_\mu\bar\mu\gamma^\mu u^c+ g_{a17} V^{\sfrac{-1}{3}}_\mu\bar\mu\gamma^\mu\gamma^5u^c+{\rm h.c.}.
\label{leptoquarks5}
\eeq
This candidate, like all vector leptoquarks, contributes to the anomalous moment through both graphs in Figures \ref{fig:vector1} and \ref{fig:vector2}.

Moving on to leptoquarks of charge \sfrac{2}{3}, we find interactions of the form
\beq
{\cal L}\supset g_{v18} V^{\sfrac{2}{3}}_\mu\bar d\gamma^\mu\mu+ g_{a18} V^{\sfrac{2}{3}}_\mu\bar d\gamma^\mu\gamma^5\mu+{\rm h.c.}.
\label{leptoquarks6}
\eeq
Candidates of charge \sfrac{4}{3} interact through
\beq
{\cal L}\supset g_{v19} V_\mu^{\sfrac{-4}{3}}\bar\mu\gamma^\mu d^c+ g_{a19}V_\mu^{\sfrac{-4}{3}}\bar\mu\gamma^\mu\gamma^5d^c+{\rm h.c.},
\label{leptoquarks7}
\eeq
and the interactions of leptoquarks with charge \sfrac{5}{3} are given by
\beq
{\cal L}\supset g_{v20} V_\mu^{\sfrac{5}{3}}\bar u\gamma^\mu\mu+ g_{a20} V_\mu^{\sfrac{5}{3}}\bar u\gamma^\mu\gamma^5\mu+{\rm h.c.}.
\label{leptoquarks8}
\eeq
\subsection{Two-Particle Contributions}

\begin{figure}
\subfigure[\label{fig:scalar1}]{\includegraphics[scale=0.3]{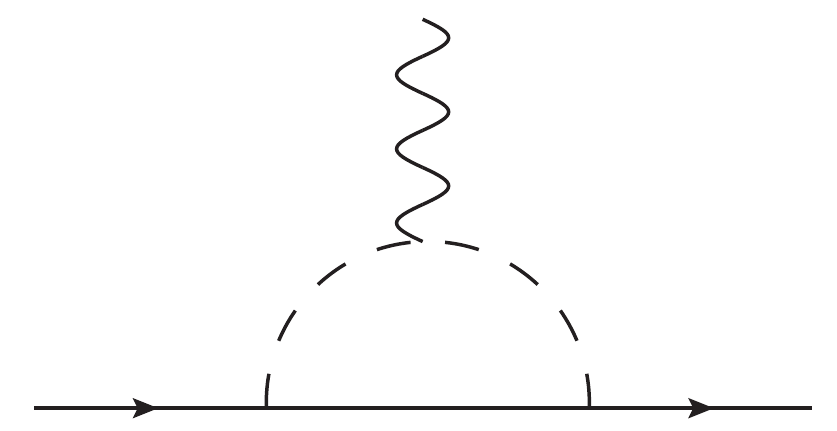}}
\subfigure[\label{fig:scalar2}]{\includegraphics[scale=0.3]{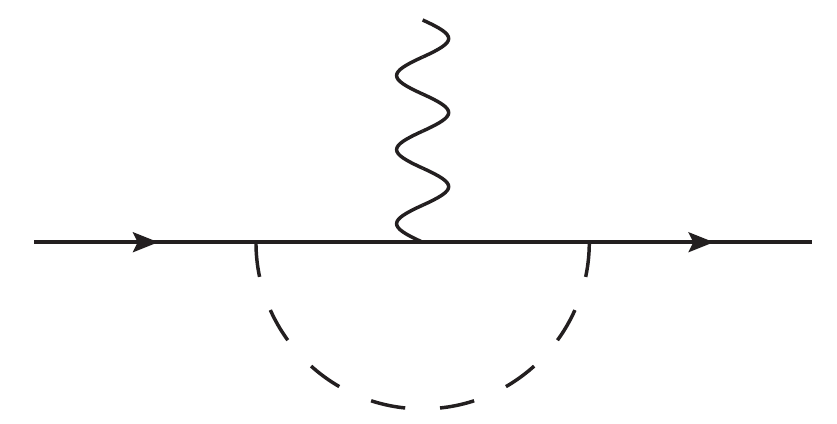}}
\subfigure[\label{fig:vector1}]{\includegraphics[scale=0.3]{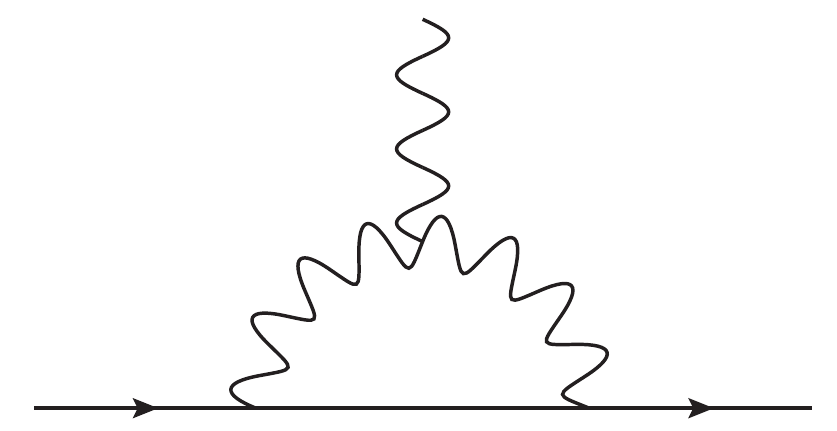}}
\subfigure[\label{fig:vector2}]{\includegraphics[scale=0.3]{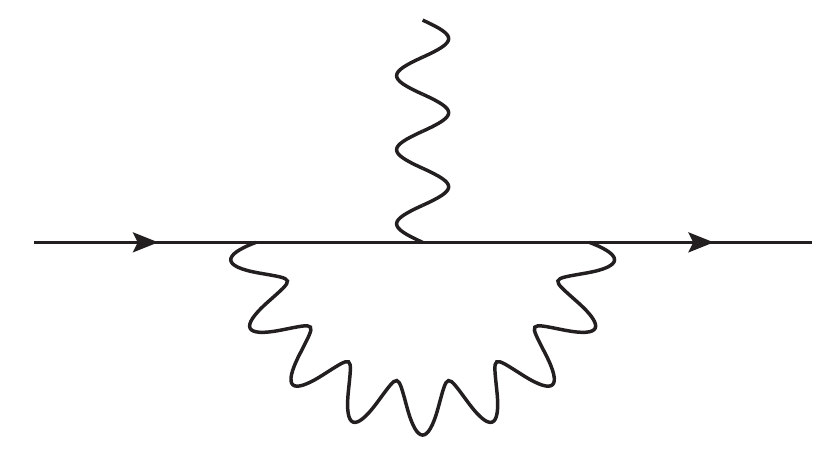}}
\caption{Feynman graphs of one-loop contributions to $a_{\mu}$. }
\end{figure}

To consider the new physics contributions which are due to two particles running in the loop we address two simplified models inspired by supersymmetric models, though other choices are certainly possible. We consider first the case of a charged scalar and new neutral fermion, contributing through Fig.\ref{fig:scalar1}, with interactions
\beq
{\cal L}\supset g_{s21} \phi^+ \bar{N} \mu + g_{a21} \phi^+ \bar{N}\gamma^5 \mu.
\label{twopar1}
\eeq
Secondly, we consider interactions of a model with a new neutral scalar and a charged fermion, contributing through Fig.\ref{fig:scalar2}, interacting as
\beq
{\cal L}\supset g_{s22} \phi \bar{\mu} E + g_{a22} \phi \bar{\mu}\gamma^5 E.
\label{twopar2}
\eeq
We have maintained the notation described above as though the new fermions carried lepton number, but it should be perfectly clear that the usual SUSY case of neutralinos, charginos, sleptons, and sneutrinos are equivalent. The reader will find that these, as well as all other, two-particle contributions can be derived from the master integrals we will present below.

\section{Contributions to $\Delta a_\mu$ and Constraints}
\label{sec:calc}

We have discussed general Lagrangians with the presence of new scalars, vectors, and leptoquarks which can give contributions to the $(g-2)_{\mu}$. Now we present the contribution due to each of those simplified models. The results in this section will be calculated in the limit of small muon mass, but exact formulae can be found in appendix \ref{appendix}. We also provide a Mathematica notebook as supplementary material for the reader's use which calculates each of these contributions.

\subsection*{Neutral Scalar}
\label{neutralscalar}

Neutral scalars give rise to contributions through the diagram shown in Fig. \ref{fig:scalar2}. Higgs-like scalars produce negligible corrections to $(g-2)_\mu$ due to their suppressed couplings to leptons, but additional neutral scalars may have stronger couplings and induce sizeable corrections to $(g-2)_\mu$. From Eq.\ref{scalar1} we notice scalar and pseudo-scalar couplings which shift $(g-2)_\mu$ by
\begin{eqnarray}
&&
\Delta a_{\mu} (h) = \frac{1}{4\pi^2}\frac{m_\mu^2}{ M_{\phi}^2 } \int_0^1 dx \frac{g_{s1}^2 \ P_{s1}(x) + g_{p1}^2 \ P_{p1}(x) }{(1-x)(1-\lambda^2 x) +\lambda^2 x}
\label{scalarmuon1}
\end{eqnarray}where $\lambda=m_{\mu}/M_{\phi}$ and,
\begin{eqnarray}
P_{s1}(x) & = &  x^2 (2 -x),\nonumber\\
P_{p1}(x) & = &  - x^3,
\label{scalarmuon2}
\end{eqnarray}which gives us,
\begin{eqnarray}
&&
\Delta a_{\mu} (\phi) = \frac{1}{4\pi^2}\frac{m_\mu^2}{ M_{\phi}^2 }\left\lbrace  g_{s1}^2\left[  \ln \left(\frac{ M_{\phi} }{m_{\mu}} \right) -\frac{7}{12}\right] + g_{p1}^2\left[ - \ln \left(\frac{ M_{\phi} }{m_{\mu}} \right) +\frac{11}{12}\right] \right\rbrace 
\label{scalarmuon3}
\end{eqnarray}
The result in Eq.\ref{scalarmuon3} is for general neutral scalars with scalar and pseudo-scalar couplings in the regime $M_{\phi} \gg m_{\mu}$. The contribution coming from only either of the pure scalar or pseudo-scalar might be easily derived from Eq.\ref{scalarmuon3} by setting either couplings $g_{p1}$ or $g_{p1}$ to zero respectively. As we mentioned earlier we are not taking into account flavour mixing throughout this work because they give negligible corrections.
Notice from Eq.\ref{scalarmuon3} that the pure neutral scalar and pseudo-scalar setups give positive and negative contributions to $(g-2)_{\mu}$. In Figs.\ref{spin0_1}-\ref{spin0_3} we shown the contribution coming from neutral scalars in three different settings. In Fig.\ref{spin0_1} we have set $g_s1=1$ and $g_{p1}=0$. In this setup $2$~TeV neutral scalars can explain the muon anomaly. However, considering a pure pseudo-scalar coupling, i.e $g_s1=0$ and $g_{p1}=1$ the contribution is negative so it cannot address the anomaly, whereas taking  $g_s1^2=1$ and $g_{p1}^2=1$ a $\sim 300$~GeV neutral scalar is a well motivated candidate to explain the $(g-2)_{\mu}$ anomaly. We emphasize that the relative sign between the scalar and pseudoscalar couplings is irrelevant to the anomalous moment, a well-known result \cite{g2generic6} that has been overlooked in some of the literature in the past \cite{g2generic4}. A purely chiral coupling of this type is what is naively expected of simple new physics (it is the only $SU(2)_L$ invariant coupling possible), but due to the fact that the scalar and pseudoscalar couplings are separable in the $a_\mu$ calculation we have chosen that basis for our investigations. This remains true for all other calculations below. 

Note that neutral scalars are also bounded by LEP searcghes for four-lepton contact interactions. For $M_{\phi} > \sqrt{s}$ these bounds require $g/M_{\phi} <2.5 \times 10^{-4}{\rm  GeV^{-1}}$ \cite{g-2LHC}.

\subsection*{Singly Charged Scalar}
\label{singlychargedS}

Singly charged scalars are predicted in a large collection of particle physics models. In Eq.\ref{scalar2} we presented a general Lagrangian involving singly charged scalars with scalar ($g_{s1}$) and pseudo-scalar ($g_{p1}$) couplings which gives rise to the $g-2_{\mu}$ correction according to Fig.\ref{fig:scalar1},
\begin{eqnarray}
&&
\Delta a_{\mu} (\phi^+) = \frac{1}{8\pi^2}\frac{m_\mu^2}{ M_{\phi^+}^2 } \int_0^1 dx \frac{g_{s2}^2 \ P_{s2}(x) + g_{p2}^2 \ P_{p2} (x) }{\epsilon^2 \lambda^2 (1-x)(1-\epsilon^{-2} x) + x}
\label{scalarmuon4}
\end{eqnarray}where 
\begin{eqnarray}
P_{s2}(x) & = &  -x (1-x)(x+\epsilon) \nonumber\\
P_{p2}(x) & = & -x (1-x)(x-\epsilon)
\label{scalarmuon5}
\end{eqnarray}with $\epsilon = m_{\nu}/m_{\mu}$ and $\lambda= m_{\mu}/M_{\phi^+}$, which results in,
\begin{eqnarray}
&&
\Delta a_{\mu} (\phi^+) = \frac{1}{4\pi^2}\frac{m_\mu^2}{ M_{\phi^+}^2 }\left\lbrace  g_{s2}^2\left(-\frac{ m_{\nu} }{4 m_{\mu}} -\frac{1}{12}\right) + g_{p2}^2 \left( \frac{ m_{\nu} }{4m_{\mu}} -\frac{1}{12}\right) \right\rbrace 
\label{scalarmuon6}
\end{eqnarray}
Eq.\ref{scalarmuon6} holds true for the last two term of Eq.\ref{scalar2}, which has a charge conjugation matrix. Therefore Eq.\ref{scalarmuon6} is a result which can be applied to any model with a charged scalar therein. We have shown the results for $g_{s2}=1, g_{p2}=0$ in Fig.\ref{spin0_1}, $g_{s2}=0, g_{p2}=1$ in Fig.\ref{spin0_2} and $g_{s2}=1, g_{p2}=1$ in Fig.\ref{spin0_3}. We can easily conclude from those results that a singly charged scalar is not a good candidate for the $(g-2)_{\mu}$ anomaly because it either gives a negative contribution or a suppressed one. 

There are collider bounds on mass of these singly charged scalars that lie in the $\sim 100-200$~GeV range \cite{chargedscalarlimits}. In specified UV models, stronger bounds might apply as well \cite{g-2LHC}.

\subsection*{Doubly Charged Scalar}

\begin{figure}[!h]
\centering
\subfigure[\label{spin0_1}]{\includegraphics[scale=0.8]{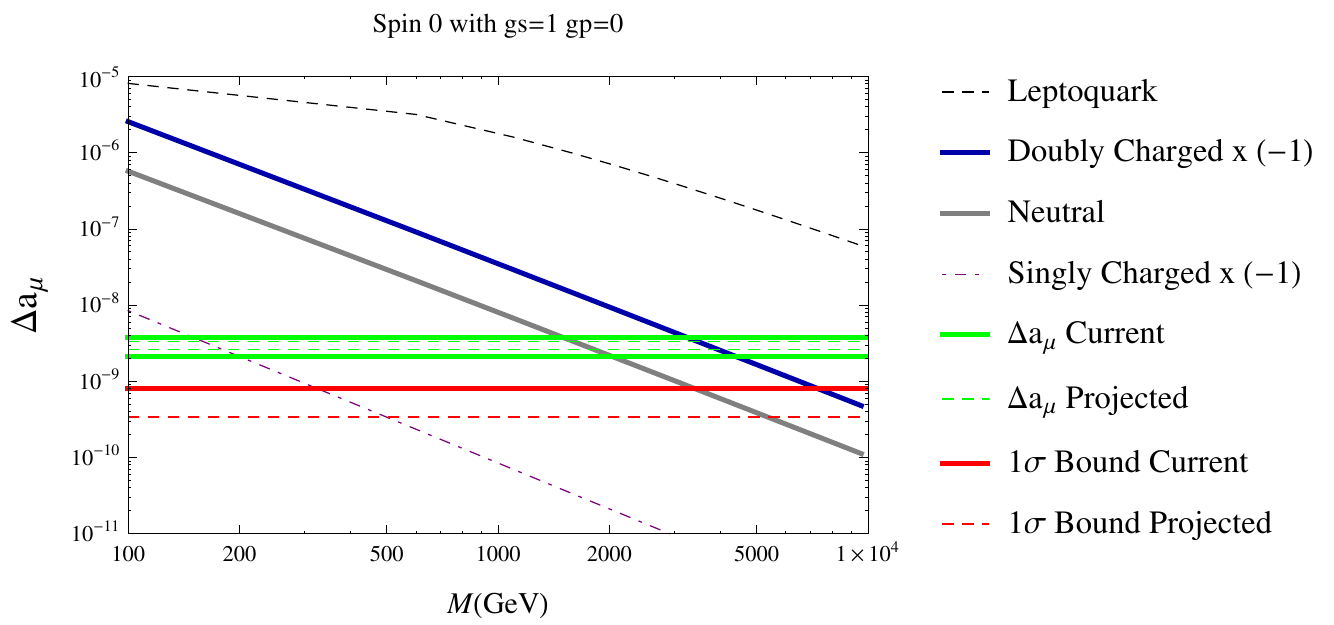}}
\subfigure[\label{spin0_2}]{\includegraphics[scale=0.8]{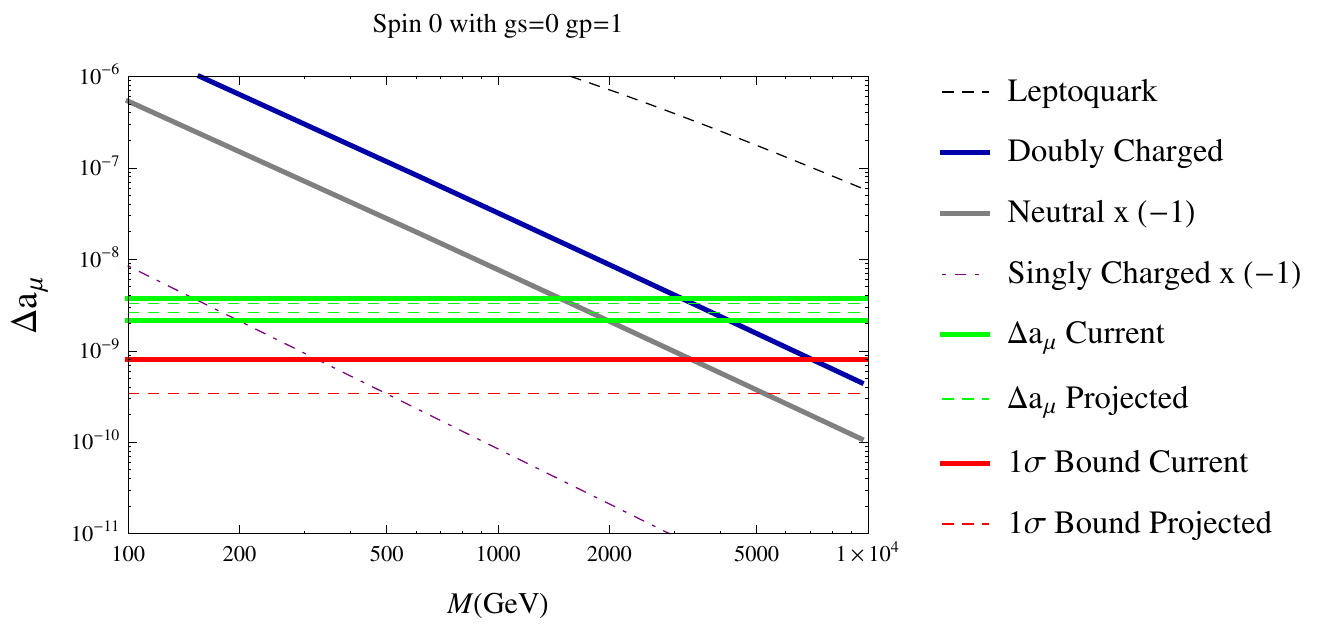}}
\subfigure[\label{spin0_3}]{\includegraphics[scale=0.8]{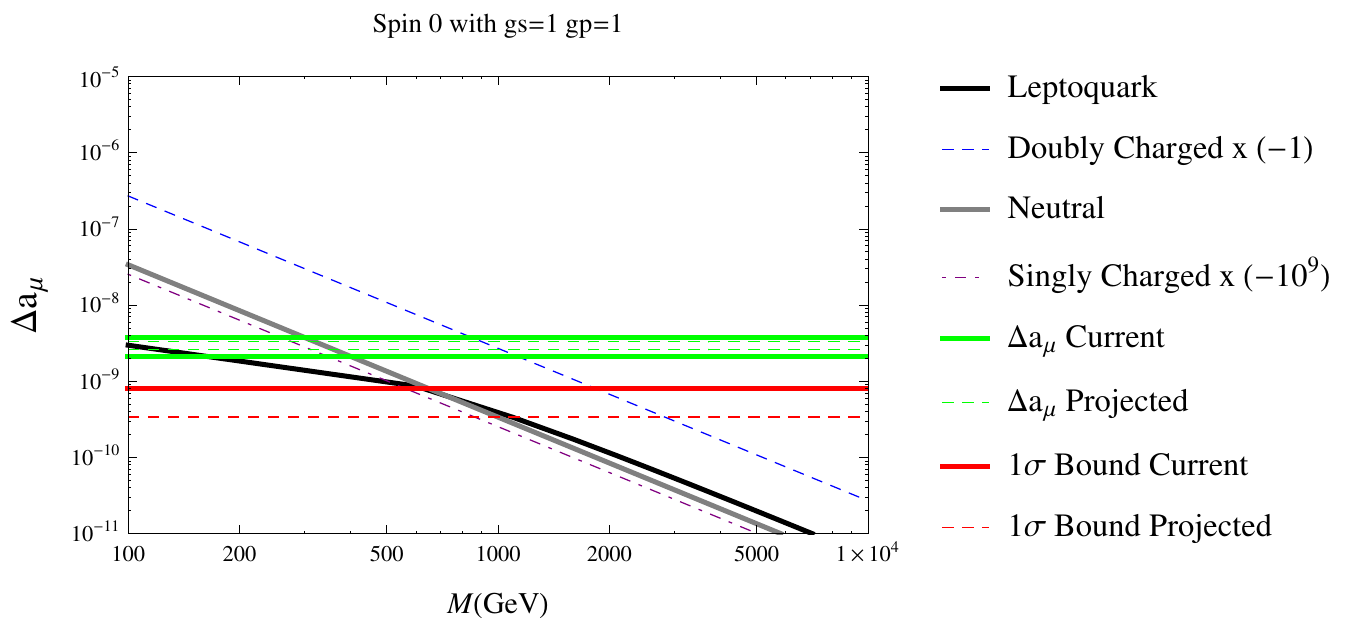}}
\caption{Contributions of Spin 0 particles to $(g-2)_{\mu}$ under three assumptions about the coupling, either pure scalar, pure pseudoscalar, or pure chiral. Keep in mind that the first two are not $SU(2)_L$ invariant without more theoretical structure, while the third is. The curves correspond as labelled in each legend to the contribution of a leptoquark, doubly-charged scalar, neutral scalar, and a singly-charged scalar. The green solid (dashed) horizontal lines are the current (projected) experimental values for $\Delta a_{\mu}$. The horizontal solid (dashed) red lines towards the bottom are the current (projected) $1\sigma$ bounds, in case the $(g-2)_{\mu}$ anomaly is resolved in a different way otherwise. Note that some of these contributions are negative or strongly suppressed; the appropriate scaling factor for each contribution is indicated in the legend of each plot.}
\end{figure}

In Eq.\ref{scalar3} we have written a general Lagrangian for the doubly charged scalar including scalar ($g_{s3}$) and pseudo-scalar ($g_{p3}$) couplings. Doubly-charged scalars are typically invoked in models with triplet of scalars \cite{doublychargedmodels1,doublychargedmodels2,doublychargedmodels3,doublychargedmodels4,doublychargedmodels5,doublychargedmodels6,doublychargedmodels7,doublychargedmodels9,doublychargedmodels10,doublychargedmodels11} and there are two diagrams contributing to the $(g-2)_\mu$ as shown in Fig.\ref{fig:scalar1}-\ref{fig:scalar2}. The contribution from each diagram is given respectively by \cite{g2generic6},
\begin{eqnarray}
\Delta a_{\mu} (\phi^{\pm \pm}) & = & (4)\times \frac{-q_H}{8 \pi^2} \left( \frac{m_\mu}{M_{\phi^{\pm \pm}}}\right)^2 \int^1_0 dx \frac{g_{s4}^2 P_s (x) + g_{p4}^2 P_p(x)}{ \lambda^2 x^2 + (1-2\lambda^2)x + \lambda^2 }+\nonumber\\
&  & (4)\times \frac{-q_f}{8 \pi^2} \left( \frac{ m_\mu}{M_{\phi^{\pm \pm}}}\right)^2 \int^1_0 dx \frac{g_{s4}^2 P^{\prime}_s (x) + g_{p4}^2 P^{\prime}_p(x) }{ \lambda^2 x^2 + (1-x)}
\label{scalarmuon7}
\end{eqnarray}where
\begin{eqnarray}
P_{s4} (x) & = x^3-x\,\,\,\,; P^{\prime}_s &=  2x^2- x^3 \nonumber\\
P_{p4} (x) & = x^3 -2x^2 +x\,\,\,\,; P^{\prime}_p & =  - x^3 
\label{scalarmuon8}
\end{eqnarray}and $\lambda = m_{\mu}/M_{\phi^{++}}$, $q_H=-2$ is the electric charge of the doubly charged scalar running in the loop, and $q_f=1$ is the electric charge of the muon in the loop. The factor of four in Eq.(\ref{scalarmuon7}) is a symmetry factor due to the presence of two identical fields in the interaction term. This expression simplifies to,
\begin{equation}
\Delta a_{\mu}(\phi^{++})= \frac{-2}{3} \left(\frac{g_{s4}^2 m_{\mu} }{\pi M_{\phi^{\pm \pm}} } \right)^2
\label{scalarmuon9}
\end{equation}when $g_{p4}=\pm g_{s4}$ and $M_{\phi^{\pm \pm}} \gg m_{\mu}$.
In the setup where either of both conditions above fail the integral in Eq.\ref{scalarmuon7} is most easily solved numerically. We have shown the results for $g_{s2}=1, g_{p2}=0$ in Fig.\ref{spin0_1}, $g_{s2}=0, g_{p2}=1$ in Fig.\ref{spin0_2} and $g_{s2}=1, g_{p2}=1$ in Fig.\ref{spin0_3}. In the first and last cases a negative contribution is found, whereas in the scenario where $g_{s2}=0$ a sizeable and positive one is obtained, showing that a $4$~TeV doubly charged scalar with pure pseudo-scalar couplings (or suppressed scalar couplings) can accommodate the muon anomaly.

Collider searches for doubly charged scalars have been explored in multiple specified models \cite{dchargedscalarlimits}, particularly including a similar simplified model approach \cite{g2models3}.

\subsection*{Neutral Lepton}

New neutral leptons are predicted in many particle physics models \cite{NeutralLepmodels1,NeutralLepmodels2,NeutralLepmodels4,NeutralLepmodels5,NeutralLepmodels6}. They also might give rise to sizeable contributions to the muon
anomalous magnetic moment according to,
\begin{eqnarray}
&&
\Delta a_{\mu} (N) = \frac{1}{8\pi^2}\frac{m_\mu^2}{ M_{N}^2 } \int_0^1 dx \frac{g_{v5}^2 \ P_{v5}(x) + g_{a5}^2 \ P_{a5} (x) }{\epsilon^2 \lambda^2 (1-x)(1-\epsilon^{-2} x) + x}
\label{leptonmuon1}
\end{eqnarray}where 
\begin{eqnarray}
P_{v5}(x) & = &  2x^2(1+x-2\epsilon)+\lambda^2(1-\epsilon)^2x(1-x)(x+\epsilon) \nonumber\\
P_{a5}(x) & = &  2x^2(1+x+2\epsilon)+\lambda^2(1+\epsilon)^2x(1-x)(x-\epsilon)
\label{leptonmuon2}
\end{eqnarray}with $\epsilon = M_{N}/m_{\mu}$ and $\lambda= m_{\mu}/M_{N}$ which results in,
\begin{eqnarray}
&&
\Delta a_{\mu} (N) = \frac{1}{4\pi^2}\frac{m_\mu^2}{ M_{N}^2 }\left\lbrace  g_{v5}^2\left[-\frac{ M_{N} }{m_{\mu}} +\frac{5}{6}\right] + g_{a5}^2 \left[ \frac{ M_N}{m_{\mu}} +\frac{5}{6}\right] \right\rbrace,
\label{leptonmuon3}
\end{eqnarray}in the $M_{W} \gg M_N$ limit. More general results are presented in Eq.(\ref{full4}) in the Appendix, and can be calculated using the supplementary Mathematica notebook available online. It is important to note, however, that a neutral lepton can give either a negative or positive contribution to $(g-2)_{\mu}$. Additionally their contribution increases with their mass when they are significantly heavier than the $W$, as seen in Figs.\ref{spin12_1}-\ref{spin12_3}. Usually such neutral leptons couple to muons through the another heavy particle such as a charged Higgs, which suppresses contributions to $a_\mu$ due to the presence of a heavy boson in the loop. Couplings to the $W$ boson are generically suppressed by mixings with SM leptons which generically get smaller with increasing mass, countering the increasing contribution we see here when we have assumed a fixed coupling. We emphasize again that these results are presented mainly to allow simple recasting to cover any model of the reader's interest, which has motivated our choice of constant couplings. 

Neutral lepton searches have been performed using LEP data imposing $M_N > 40$~GeV \cite{chargedscalarlimits}. In the case where the new neutral fermion has interactions identical to those of the standard model neutrinos the bounds reach 2.4TeV \cite{chargedscalarlimits}. Complementary bounds have been found in Ref.\cite{g-2LHC}.

\subsection*{Charged Lepton}

\begin{figure}[!h]
\centering
\subfigure[\label{spin12_1}]{\includegraphics[scale=0.8]{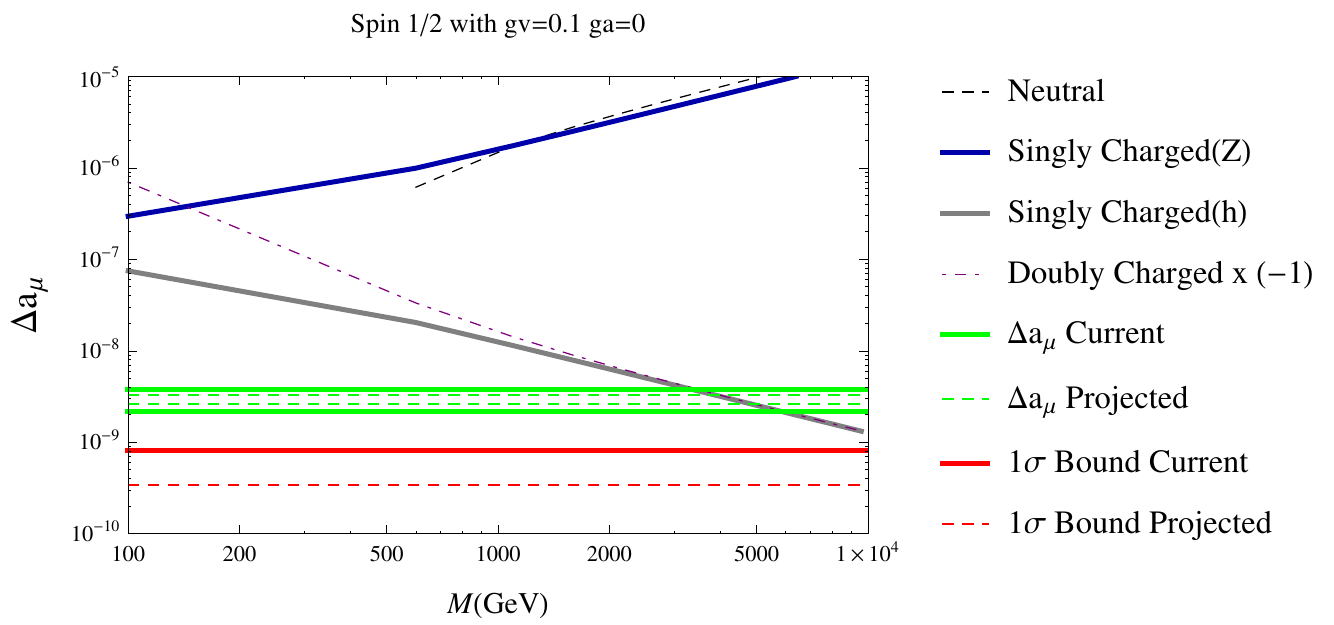}}
\subfigure[\label{spin12_2}]{\includegraphics[scale=0.8]{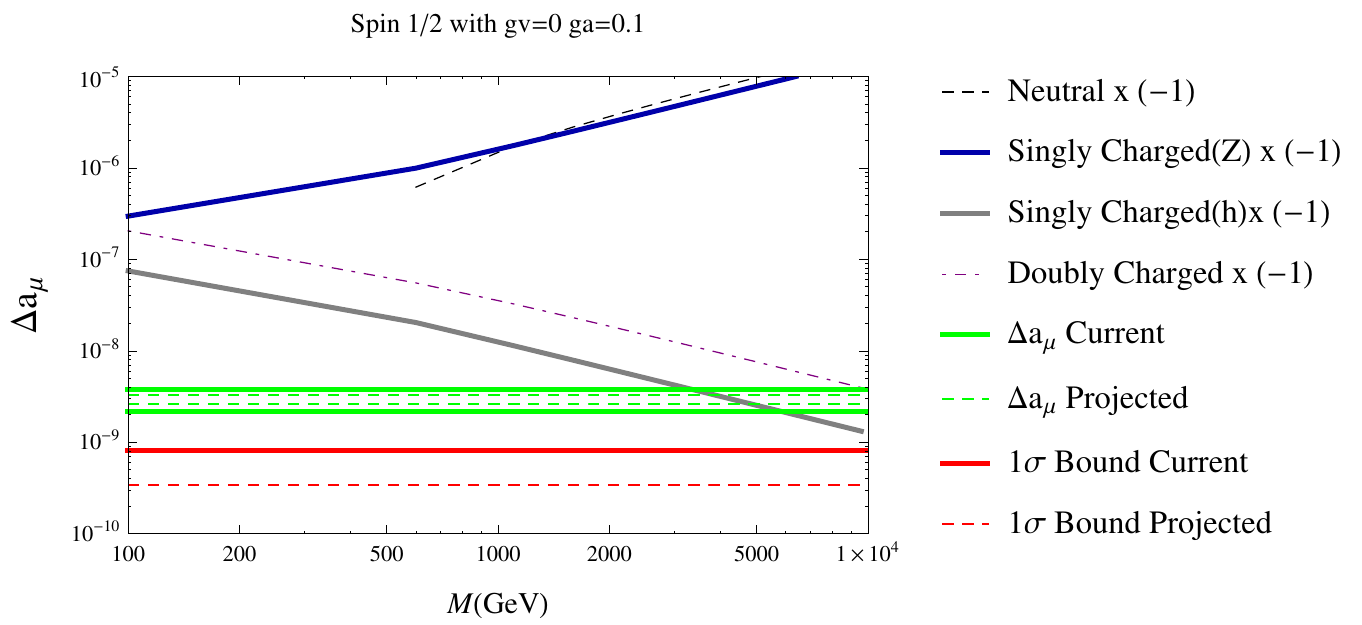}}
\subfigure[\label{spin12_3}]{\includegraphics[scale=0.8]{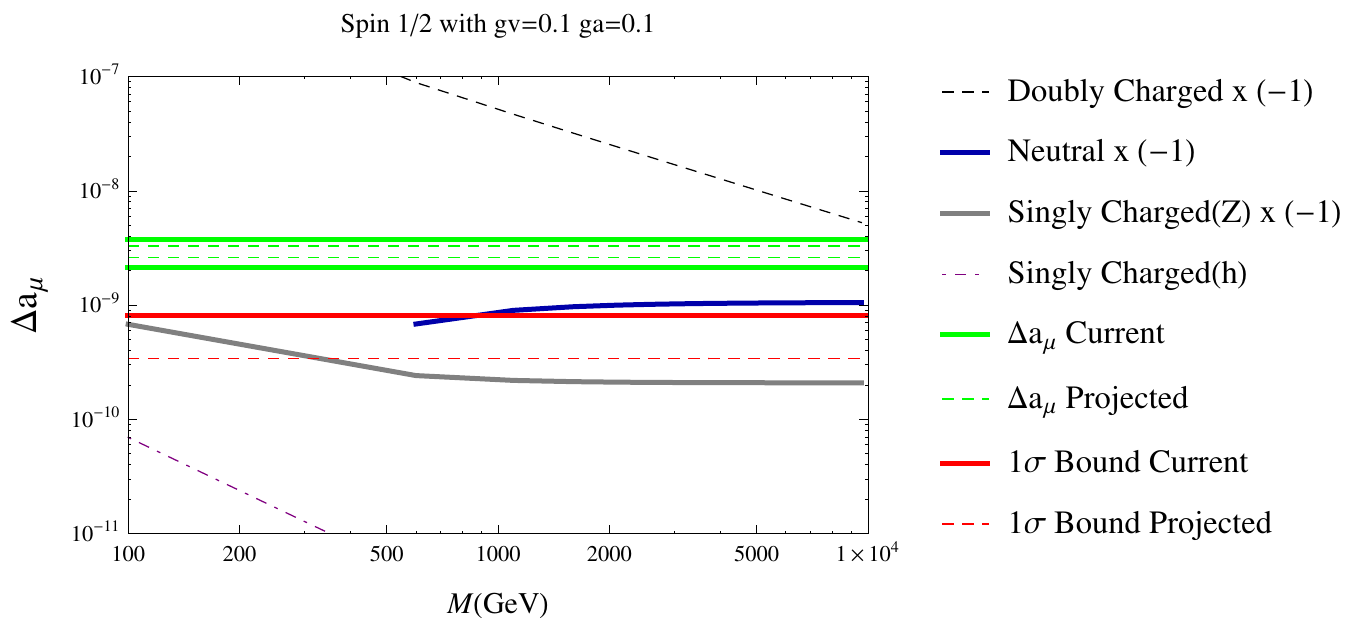}}
\caption{Contributions of Spin 1/2 particles to the muon anomalous magnetic moment assuming either a pure vector, pure axial, or pure chiral coupling. Again note that only pure chiral couplings are, without further theoretical structure, $SU(2)_L$ invariant. The curves give the contributions due to a new neutral lepton, new charged lepton coupled to the $Z$ boson and Higgs boson, and a doubly charged lepton, as indicated in the figure legends. The green solid (dashed) horizontal lines are the current (projected) experimental values for $\Delta a_{\mu}$. The horizontal solid (dashed) red lines towards the bottom are the current (projected) $1\sigma$ bounds, assuming the $(g-2)_{\mu}$ anomaly is resolved without this new physics. Note that some contributions are negative, and the appropriate scaling factors are given in the plot legends.}
\end{figure}

Multiple models \cite{chargedLepmodels1,chargedLepmodels2,chargedLepmodels3,chargedLepmodels4,chargedLepmodels5} predict the existence of new charged leptons that give sizeable contributions to the $(g-2)_\mu$ through $Z$ and Higgs couplings discussed in Eq.\ref{lepton2}. Here we present the results for these two possibilities separately.
\begin{itemize}
\item $Z$-Mediated

This process is exhibited in Fig.\ref{fig:vector1}. The corresponding integral is given by,
\begin{eqnarray}
&&
\Delta a_{\mu} (E) = \frac{1}{8\pi^2}\frac{m_\mu^2}{ M_{Z}^2 } \int_0^1 dx \frac{g_{v6}^2 \ P_{v6}(x) + g_{a6}^2 \ P_{a6} (x) }{(1-x)(1-\lambda^{2} x) +\epsilon^2 \lambda^2 x}
\label{leptonmuon4}
\end{eqnarray}where 
\begin{eqnarray}
P_{v6}(x) & = &  2x(1-x)(x-2(1-\epsilon))+\lambda^2(1-\epsilon)^2x^2(1+\epsilon-x) \nonumber\\
P_{a6}(x) & = &  2x^2(1+x+2\epsilon)+\lambda^2(1+\epsilon)^2x(1-x)(x-\epsilon)
\label{leptonmuon5}
\end{eqnarray}with $\epsilon = M_{E}/m_{\mu}$ and $\lambda= m_{\mu}/M_{Z}$.
Therefore the contribution of a generic singly charged lepton Z-mediated to the muon anomalous magnetic moment is given by,
\begin{eqnarray}
&&
\Delta a_{\mu} (E) = \frac{1}{4\pi^2}\frac{m_\mu^2}{ M_{Z}^2 }\left\lbrace  g_{v6}^2\left[\frac{ M_{E} }{m_{\mu}} -\frac{2}{3}\right] + g_{a6}^2 \left[ -\frac{ M_E}{m_{\mu}} -\frac{2}{3}\right] \right\rbrace,
\label{leptonmuon6}
\end{eqnarray}in the  $M_Z \gg M_{E}$ limit, otherwise one should either solve Eq.(\ref{leptonmuon4}) numerically using our Mathematica notebook or use the full analytical expression given in Eq.(\ref{full2}) in the Appendix. Of course, there is no particular requirement that the Z boson mediate this process. As the Z mass and couplings are free parameters in our computation, one can easily replace them by any other neutral boson mass and couplings. In other words, the integral in Eq.(\ref{leptonmuon4}) can be straightforwardly applied to processes where a new $Z^{\prime}$ boson is the mediator of the charged lepton contribution to the muon anomalous magnetic moment.  

The results for the charged lepton can be found in Figs.\ref{spin12_1}-\ref{spin12_3}. As for the neutral lepton case, the contribution increases with the mass of the lepton in agreement with Ref\cite{g2generic4}. From Fig.\ref{spin12_1} we conclude that pure vector couplings with unit strength of the $Z$ boson to muons and the new lepton induce too large a positive contribution to the muon magnetic moment, prohibiting such strong interactions. Interactions of this type with strength $g_{v6}=0.1$ rather than 1 give an approximately correct result for 100 GeV new leptons, and the coupling must be smaller yet as the mass of the new lepton increases. Furthermore, negative contributions are found in the pure vector-axial case, as shown in Fig.\ref{spin12_2}. For $g_{v6}=g_{a6}$ negative corrections are obtained, as exhibited in Fig.\ref{spin12_3}.

\item Higgs-mediated

A neutral scalar boson can also mediate the charged lepton contribution to $(g-2)_{\mu}$ through the diagram shown in Fig.\ref{fig:scalar2}. The contribution is determined by,
\begin{eqnarray}
&&
\Delta a_{\mu} (E) = \frac{1}{8\pi^2}\frac{m_\mu^2}{ M_{h}^2 } \int_0^1 dx \frac{g_{s7}^2 \ P_{s7}(x) + g_{p7}^2 \ P_{p7} (x) }{(1-x)(1-\lambda^{2} x) +\epsilon^2 \lambda^2 x}
\label{leptonmuon7}
\end{eqnarray}where 
\begin{eqnarray}
P_{s7}(x) & = &  x^2 (1+\epsilon -x) \nonumber\\
P_{p7}(x) & = &  x^2 (1-\epsilon -x)
\label{leptonmuon8}
\end{eqnarray}with $\epsilon = M_{E}/m_{\mu}$ and $\lambda= m_{\mu}/M_{h}$. In the limit $M_h \gg M_E$ it simplifies to,
\begin{eqnarray}
&&
\Delta a_{\mu} (E) = \frac{1}{4\pi^2} \frac{m_{\mu}^2}{ M_{h}^2 }\left\lbrace  g_{s7}^2\left[ \frac{ M_{E} }{m_{\mu}}\left( \ln\left( \frac{M_E}{m_{\mu}}\right) -\frac{3}{4} \right) +\frac{1}{6} \right] +  g_{p7}^2\left[ -\frac{ M_{E} }{m_{\mu}}\left( \ln\left(\frac{M_E}{m_{\mu}}\right) -\frac{3}{4} \right) +\frac{1}{6} \right] \right\rbrace \nonumber\\
\label{leptonmuon9}
\end{eqnarray}
The result for any coupling and mass regime can be found either using the full analytical formula given by Eq.\ref{full1} or numerically using our Mathematica code. We have shown the results for $g_{s7}=1, g_{p7}=0$ in Fig.\ref{spin12_1}, $g_{s7}=0, g_{p7}=1$ in Fig.\ref{spin12_2} and $g_{s7}=1, g_{p7}=1$ in Fig.\ref{spin12_3}.From Figs.\ref{spin12_1}-\ref{spin12_3} it is clear that only the pure scalar coupling scenario is able to address the muon anomaly, and that requires $4$~TeV charged lepton. We have used $M_h=125$~GeV in this result. We will investigate the regime where a heavy Higgs replaces the SM Higgs further below. 

Note that the L3 Collaboration has placed a limit $M_E > 100$~GeV for a fourth generation of leptons \cite{chargedleptonbound}. Other existing bound on charged lepton can be found in Ref.\cite{chargedlepboundothers}.

\end{itemize}
\subsection*{Doubly Charged Lepton}

There are many models where new multiplets of leptons are predicted, and these can include exotic
doubly charged leptons \cite{doublyCLeptons1,doublyCLeptons2,doublyCLeptons3}.
The contribution to the muon anomalous magnetic moment of these exotic
leptons, which are exhibited in Figs.\ref{fig:vector1}-\ref{fig:vector2} is
\begin{eqnarray}
& \Delta a_{\mu} (\psi) =& \frac{1}{8\pi^2}\frac{m_\mu^2}{ M_{\psi}^2 } \int_0^1 dx \frac{g_{v8}^2 \ P_{v8}(x) + g_{a8}^2 \ P_{a8} (x) }{ \epsilon^2 \lambda^2 (1-x)(1-\epsilon^{-2}x)+x}+\nonumber\\
&       &\frac{1}{8\pi^2} \frac{Q_{\psi}m_\mu^2}{ M_{\psi}^2 } \int_0^1 dx \frac{g_{v8}^2 \ P_{v8}^{\prime}(x) + g_{a8}^2 \ P_{a8}^{\prime} (x) }{ (1-x)(1-\lambda^{2}x)+\epsilon^2 \lambda^2 x},
\label{leptonmuon10}
\end{eqnarray}where 
\begin{eqnarray}
P_{v8}(x) & = &  2x^2(1+x-2\epsilon)+\lambda^2(1-\epsilon)^2x(1-x)(x+\epsilon)\nonumber\\
P_{a8}(x) & = &  2x^2(1+x+2\epsilon)+\lambda^2(1+\epsilon)^2x(1-x)(x-\epsilon)\nonumber\\
P_{v8}^{\prime}(x) & = &  2x(1-x)\left(x-2(1-\epsilon) \right)+\lambda^2(1-\epsilon)^2x^2(1+\epsilon-x)\nonumber\\
P_{a8}^{\prime}(x) & = &  2x(1-x)\left(x-2(1+\epsilon) \right)+\lambda^2(1+\epsilon)^2x^2(1-\epsilon-x)\nonumber\\,
\label{leptonmuon11}
\end{eqnarray}with $Q_{\psi}=-2$, $\epsilon = M_{\psi}/m_{\mu}$ and $\lambda= m_{\mu}/M_{W}$, which in the limit $M_{\psi} \gg m_{\mu},M_{Z}$ gives
\begin{eqnarray}
&\Delta a_{\mu} (\psi) =& \frac{1}{4\pi^2}\frac{m_\mu^2}{ M_{\psi}^2 }\left\lbrace  g_{v8}^2\left[-\frac{ M_{\psi} }{m_{\mu}} +\frac{5}{6}\right] + g_{a8}^2 \left[ \frac{ M_{\psi}}{m_{\mu}} +\frac{5}{6}\right] \right\rbrace\nonumber\\
&  & +\frac{Q_{\psi} m_{\mu}^2}{4 \pi^2 M_W^{2}}\left(\frac{g^2_{v8}}{3} - \frac{5g^2_{a8}}{3}\right).
\label{leptonmuon12}
\end{eqnarray}
As for the result for any mass regime, the reader can make use of Eq.\ref{full2} and Eq.\ref{full4} and find full analytical expressions or evaluate the integral numerically. Doubly Charged leptons give uniformly negative contributions to the muon magnetic moment and hence they cannot accommodate the anomaly.

Collider search ounds on such doubly charged leptons can be found in Refs.\cite{Dleptonbound} and they are in the $\sim 100$~GeV range.

\subsection*{Neutral Vector}

Now let us consider the contribution of the new neutral gauge boson, which we denote as $Z'$. The only diagram which appears with this particle is exhibited in Figure \ref{fig:vector2}. The result is given by \cite{g2generic4} to be
\begin{eqnarray}
&&
\Delta a_{\mu} (Z^{\prime}) = \frac{m_\mu^2}{8\pi^2 M_Z^{\prime 2} }\int_0^1 dx \frac{g^2_{v9} P_{v9}(x)+ g^2_{a9} P_{a9}(x) }{(1-x)(1-\lambda^2 x) +\lambda^2 x},
\label{vectormuon1}
\end{eqnarray}, where $\lambda = m_{\mu}/M_{Z^{\prime}}$ and
\begin{eqnarray}
P_{v9}(x) & = & 2 x^2 (1-x) \nonumber\\
P_{a9}(x) & = & 2 x(1-x)\cdot (x-4)- 4\lambda^2 \cdot x^3.
\label{vectormuon2}
\end{eqnarray}
These integrals simplify to give a contribution of 
\begin{equation}
\Delta a_{\mu}(Z^{\prime}) = \frac{m_{\mu}^2}{4 \pi^2 M_Z^{\prime 2}}\left(\frac{1}{3}g^2_{v9} - \frac{5}{3}g^2_{a9}\right)
\label{vectormuon3}
\end{equation}in the limit $M_{Z^{\prime}} \gg m_{\mu}$.
This is the contribution of the $Z^{\prime}$ to the muon anomaly magnetic moment. Notice that, depending on the values of the vector and axial couplings, the contribution can be either positive or negative. From Figs.\ref{spin1_1}-\ref{spin1_3} we conclude that pure vector or axial neutral vectors with $M_{Z^{\prime}}< 10$~TeV and order one couplings are excluded. However, when both couplings are unit strength, a $\sim 1$~TeV $Z^{\prime}$ naturally addresses the anomaly. 

From LEP measurements a 95\% C.L upper bound might apply for $g_{v9}=g_{a9}$ and $M_Z^{\prime} > \sqrt{s}$ that reads $g_{v9}/M_Z^{\prime} < 2.2\times 10^{-4}{\rm GeV^{-1}}$, ruling out the possibility of a single $Z^{\prime}$ boson to explain the anomaly in agreement with \cite{g-2LHC}. Additional bounds are present in the literature \cite{g-2LHC,chargedscalarlimits}.

\subsection*{Singly Charged Vector} 

Singly charged vectors are predicted in many extended gauge theories \cite{chargedVec1,chargedVec2,chargedVec3,chargedVec4}. Their contributions
to $(g-2)_{\mu}$ show up in the form of Fig.\ref{fig:vector1} and it is determined by Eq.(\ref{vectormuon4}) as follows, 
\begin{eqnarray}
&&
\Delta a_{\mu} (W^{\prime}) = \frac{1}{8\pi^2}\frac{m_\mu^2}{ M_{V^+}^2 } \int_0^1 dx \frac{g_{v10}^2 \ P_{v10}(x) + g_{a10}^2 \ P_{a10} (x) }{\epsilon^2 \lambda^2 (1-x)(1-\epsilon^{-2} x) + x},
\label{vectormuon4}
\end{eqnarray}where 
\begin{eqnarray}
P_{v10}(x) & = &  2x^2(1+x-2\epsilon)+\lambda^2(1-\epsilon)^2 x(1-x)(x+\epsilon) \nonumber\\
P_{a10}(x) & = &  2x^2(1+x+2\epsilon)+\lambda^2(1+\epsilon)^2 x(1-x)(x-\epsilon),
\label{vectormuon5}
\end{eqnarray}with $\epsilon = m_{\nu}/m_{\mu}$ and $\lambda= m_{\mu}/M_{W^{\prime}}$. This simplifies to
\begin{eqnarray}
&&
\Delta a_{\mu} (W^{\prime}) = \frac{1}{4\pi^2}\frac{m_\mu^2}{ M_{W^{\prime}}^2 } \left[g_{v10}^2 \left( \frac{5}{6} - \frac{m_{\nu}}{m_{\mu}}\right)  + g_{a10}^2 \left(  \frac{5}{6} + \frac{m_{\nu}}{m_{\mu}} \right) \right]
\label{vectormuon6}
\end{eqnarray}in the regime $M_{W^{\prime}} \gg m_{\mu}$. One can clearly see that a singly charged vector boson rises as a natural candidate to explain the $(g-2)_{\mu}$ anomaly because it gives always positive contributions and for couplings of order one as we expect from gauge couplings, singly charged vector with masses of $\sim1$~TeV might account for the anomaly. These contributions are plotted in Figs.\ref{spin1_1}-\ref{spin1_3}. 

Searches in the regime where this new charged boson interacts only with right handed neutrinos, i.e when $g_{a10}=-g_{v10}$ give a 95\% C.L bound from LEP using effective operators which reads $g_{v10}/M_{W^{\prime}} < 4.8 \times 10^{-3} {\rm GeV^{-1}}$ \cite{g-2LHC}.

\subsection*{Doubly Charged Vector} 
 
\begin{figure}[!h]
\centering
\subfigure[\label{spin1_1}]{\includegraphics[scale=0.9]{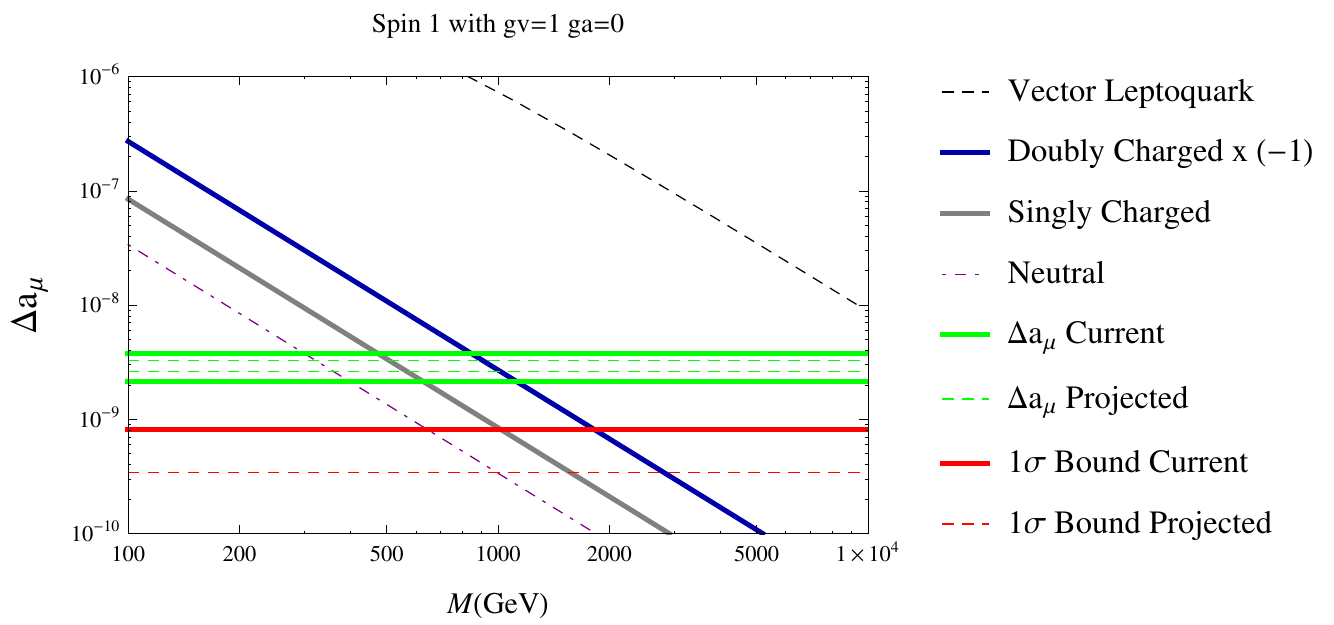}}
\subfigure[\label{spin1_2}]{\includegraphics[scale=0.9]{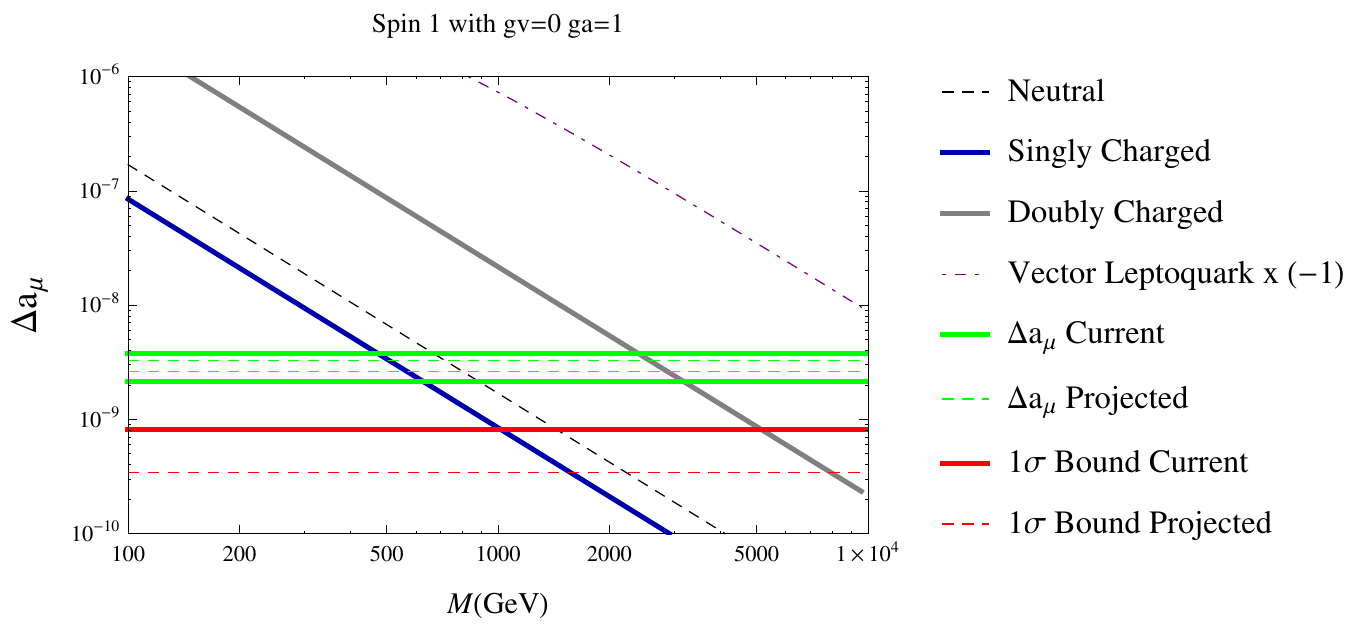}}
\subfigure[\label{spin1_3}]{\includegraphics[scale=0.9]{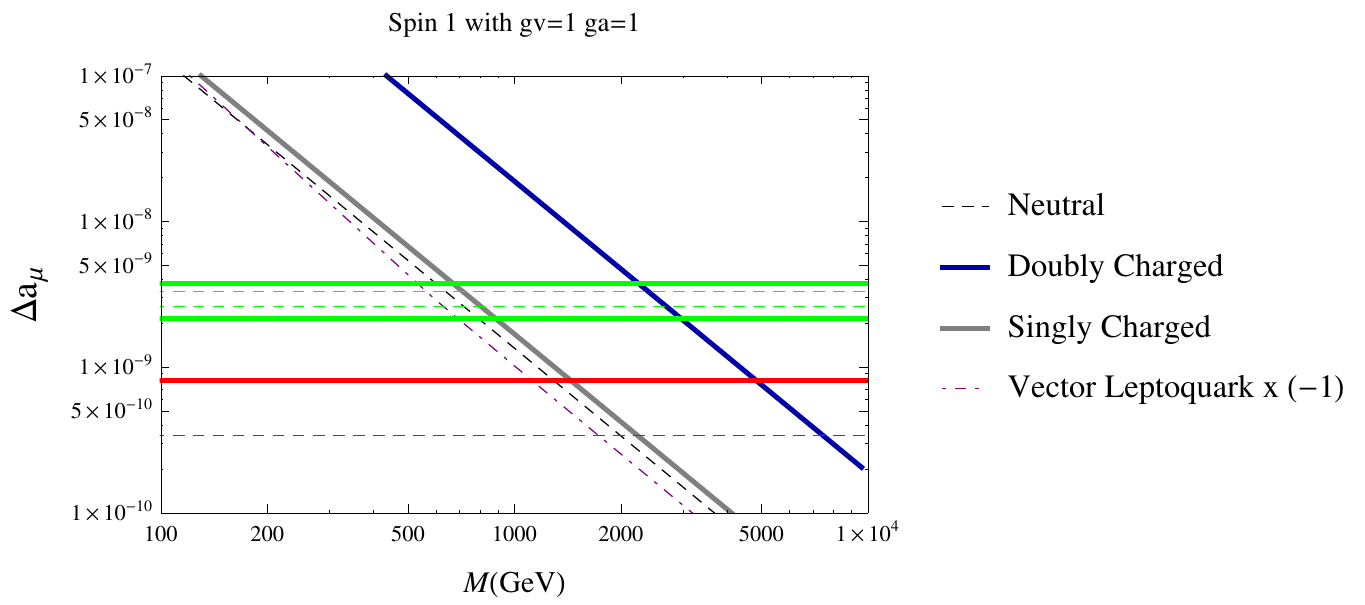}}
\caption{Contributions of spin 1 particles to $a_\mu$. Keep in mind that the purely vector and purely axial coupling cases are not consistent with $SU(2)_L$ without additional theoretical structure. The curves correspond to neutral, charged, doubly-charged, and leptoquark vector bosons, as indicated in the plot legends. The green solid (dashed) horizontal lines are the current (projected) experimental values for $\Delta a_{\mu}$. The horizontal solid (dashed) red lines towards the bottom are the current (projected) $1\sigma$ bounds, in case the $(g-2)_{\mu}$ anomaly is resolved in a different way otherwise. Note that some contributions are negative, and the appropriate scaling factors are given in the plot legends for those cases.}
\end{figure}

The doubly-charged vector boson, similarly to the doubly-charged scalar, gives rise to two diagrams that contribute to the $(g-2)_\mu$. The first one, shown in Fig. \ref{fig:vector1}, is similar to the singly-charged gauge boson, with two differences: a multiplying factor of 4 due to the symmetry factors arising from identical fields in the interaction term, and an additional factor of 2 arising from the larger charge of the boson \cite{g2generic6}.
The second diagram, shown in Fig. \ref{fig:vector2}, is similar to the $Z^{\prime}$ one, but we once again have a factor of 4 due to the identical fields, and we also have a relative negative sign due to the opposite charge of the muon running in the loop. Hence we find,
\begin{eqnarray}
&\Delta a_{\mu} (U^{\pm \pm}) &=8\times \frac{1}{8\pi^2}\left( \frac{m_\mu}{M_{U^{\pm \pm}}}\right)^2\int_0^1 dx \frac{g_{v11}^2 P_{v11}(x) + g_{a11}^2 P_{a11} (x) }{\lambda^2(1-x)^2 + x }\nonumber\\
&  & (-4)\times \frac{1}{8\pi^2}\left( \frac{m_\mu}{M_{U^{\pm \pm}}}\right)^2\int_0^1 dx \frac{g^2_{v12} P_{v12}^{\prime}(x)+ g^2_{a12} P_{a12}^{\prime} (x) }{(1-x)(1-\lambda^2 x) + \lambda^2 x},
\label{Vcontri}
\end{eqnarray}where $\lambda = m_{\mu}/M_{U^{\pm \pm}}$, and
\begin{eqnarray}
P_{v12}(x) & = & 2 x^2(x-1) \nonumber\\
P_{a12}(x) & = & 2 x^2(x+3)+4 \lambda^2 \cdot x (1-x)(x-1), \nonumber\\
P_{v12}^{\prime}(x) & = & 2 x (1-x)\cdot x \nonumber\\
P_{a12}^{\prime}(x) & = & 2 x(1-x)\cdot (x-4)- 4\lambda^2 \cdot x^3.
\end{eqnarray}
Hence the total doubly-charged vector contribution is given by,
\begin{eqnarray}
\Delta a_{\mu} (U^{\pm \pm})&=& \frac{ m_{\mu}^2}{\pi^2 M_{U^{\pm \pm}}^2}\left( \frac{-2}{3}g_{v12}^2 + \frac{16}{3}g_{a12}^2 \right)
\label{doublyvector}
\end{eqnarray} 
Notice that purely-vector doubly charged bosons give negative contributions to $g-2_{\mu}$, whereas the purely-axial ones are positive. Furthermore, for unit axial couplings, $1-2$~ TeV doubly-charged bosons are natural candidates to explain the $g-2_{\mu}$ anomaly as can be easily noted in Figs.\ref{spin1_1}-\ref{spin1_3}.

\subsection*{Scalar Leptoquarks}
Scalar leptoquarks are predicted in a variety of particle physics models \cite{scalarleptoquarks1,scalarleptoquarks2,scalarleptoquarks3,scalarleptoquarks4,
scalarleptoquarks5}. They contribute to the muon anomalous magnetic moment through the diagrams shown in Figs.\ref{fig:scalar1} and \ref{fig:scalar2}. We have listed in Eqs.\ref{leptoquarks1}-\ref{leptoquarks4} possible interactions involving scalar leptoquarks that give rise to corrections to $(g-2)_{\mu}$ \cite{scalarleptoquarks6}. Here we calculate all their contributions simultaneously, finding
\begin{eqnarray}
&&
\Delta a_{\mu} (\Phi) = \frac{1}{8\pi^2}\frac{N_c Q_q m_\mu^2}{ M_{\Phi}^2 } \int_0^1 dx \ \frac{g_{s13}^2 \ P_{s13}(x) + g_{p13}^2 \ P_{p13}(x) }{(1-x)(1-\lambda^2 x) +\epsilon^2 \lambda^2 x}\nonumber\\+
&&
\frac{1}{8\pi^2}\frac{N_c Q_{\Phi} m_\mu^2}{ M_{\Phi}^2 } \int_0^1 dx \ \frac{g_{s13}^2 \ P_{s13}^{\prime}(x) + g_{p13}^2 \ P_{p13}^{\prime}(x) }{\epsilon^2 \lambda^2 (1-x)(1-\epsilon^{-2}x) +x}
\label{leptoSmuon1}
\end{eqnarray}where $\epsilon=m_{q}/m_{\mu}$ and $\lambda=m_{\mu}/M_{\Phi}$, $m_{q} (Q_q)$ is the mass (electric charge) of the quark running in the loop, and
\begin{eqnarray}
P_{s13}^{\prime}(x) & = &  x^2 (1+\epsilon -x),\nonumber\\
P_{p13}^{\prime}(x) & = &  x^2 (1+\epsilon -x),\nonumber\\
P_{s13}^{\prime}(x) & = &  -x(1-x)(x+\epsilon),\nonumber\\
P_{p13}^{\prime}(x) & = &  -x(1-x)(x-\epsilon)
\label{leptoSmuon2}
\end{eqnarray}
After algebra this integral simplifies to,
\begin{eqnarray} 
\Delta a_{\mu} (\Phi) & = & - \frac{N_c m_{\mu}}{8 \pi^2 M_{\Phi}^2} \left[ m_q (g_{s13}^2 -g_{a13}^2) \left( Q_{\Phi} f_1 + Q_q f_2 \right) + 2m_{\mu} (g_{s13}^2 + g_{a13}^2) \left( Q_{\Phi} f_3 + Q_q f_4 \right) \right],\nonumber\\
\label{leptoSmuon3}
 \end{eqnarray}where $Q_{\Phi}$ and $Q_q$ are the electric charge of the leptoquark and the quark respectively, and $f_i$ are
\begin{eqnarray} 
f_1 & = & \frac{1}{2(1-x)^3}\left(1-x^2+2x\ln x\right) \nonumber\\
f_2 & = & \frac{1}{2(1-x)^3}\left(3-4x+x^2+2\ln x\right) \nonumber\\
f_3 & = & \frac{1}{12(1-x)^4}\left(-1 + 6x -3x^2 -2x^3 +6x^2 \ln x\right)\nonumber\\
f_4 & = &\frac{1}{2(1-x)^3}\left(2 + 3x -6x^2 + x^3 +6 x \ln x \right).
\label{leptoSmuon4}  
\end{eqnarray} 
For instance, when the quark in consideration is the top quark, $Q_{q}=2/3$. One can clearly see in Figs.\ref{spin0_1}-\ref{spin0_3} that a $\sim 1$~TeV scalar Leptoquark with scalar and pseudo-scalar couplings are natural candidates for the $(g-2)_{\mu}$ anomaly. Additionally we can exclude pure order one scalar and pseudo-scalar couplings for Leptoquarks lighter than $10$~TeV. Suppressed couplings are required in these cases to resolve the $(g-2)_{\mu}$ discrepancy.

We would like to point out that in addition to the $g-2_{\mu}$ anomaly there are additional bounds on leptoquark states obtained both directly and indirectly. Direct limits arise from their production cross sections at colliders, while indirect limits are calculated from the bounds on the leptoquark-induced four-fermion interactions which are 
obtained from low-energy experiments or from collider experiments below threshold. Based on the scalar leptoquarks pair production CMS has reported a stringent constraint on scalar leptoquarks requiring $M_{Phi} > 1070$~GeV for scalar leptoquarks of second generation decaying with 100\% branching ratio into $\mu q$ \cite{leptoquarkbound}.

\subsection*{Vector Leptoquarks}

Vector Leptoquarks arise in a large collection of extensions of the Standard Model (SM) such as composite models \cite{Vectorleptoquarks1}, Grand Uniﬁed Theories \cite{Vectorleptoquarks2,Vectorleptoquarks3} and E6 models \cite{Vectorleptoquarks4,Vectorleptoquarks5}. Their contributions to the $(g-2)_{\mu}$ comes from the diagrams in Figs.\ref{fig:vector1} and \ref{fig:vector2}, and are found to be
\begin{eqnarray}
\Delta a_{\mu} (V) & = & \frac{m_\mu^2 N_c Q_{q}}{8\pi^2 M_V^2 }\int_0^1 dx \frac{g^2_{v} P_{v}(x)+ g^2_{a} P_{a}(x) }{(1-x)(1-\lambda^2 x) +\epsilon^2 \lambda^2 x}+\nonumber\\
&  &\frac{m_\mu^2 N_c Q_{V}}{8\pi^2 M_V^2 }\int_0^1 dx \frac{g^2_{v} P_{v}^{\prime}(x)+ g^2_{a} P_{a}^{\prime}(x) }{\epsilon^2 \lambda^2(1-x)(1-\epsilon^{-2}x)+x},
\label{leptoVmuon1}
\end{eqnarray}with $\epsilon=m_{q}/m_{\mu}$, $\lambda = m_{\mu}/M_V$ and
\begin{eqnarray}
P_{v}(x) & = & 2x(1-x)\left(x-2(1-\epsilon)\right)+\lambda^2(1-\epsilon)^2x^2(1+\epsilon-x),\nonumber\\
P_{a}(x) & = & 2x(1-x)\left(x-2(1+\epsilon)\right)+\lambda^2(1+\epsilon)^2x^2(1-\epsilon-x),\nonumber\\
P_{v}^{\prime}(x) & = & 2x^2(1+x-2\epsilon)-\lambda^2(1-\epsilon^2)\left(-x(1-x)(x+\epsilon)\right), \nonumber\\
P_{a}^{\prime}(x) & = & 2x^2(1+x+2\epsilon)-\lambda^2(1+\epsilon^2)\left(-x(1-x)(x-\epsilon)\right).
\label{leptoVmuon2}
\end{eqnarray}
These integrals simplify to,
\begin{eqnarray}
\Delta a_{\mu} (V) & = & \frac{m_\mu^2 N_c Q_{q}}{8\pi^2 M_V^2 }\left[ g_v^2\left(\frac{-4}{3}+2\epsilon\right) +g^2_a\left(\frac{-4}{3}-2\epsilon\right) \right]\nonumber\\
& &+\frac{m_\mu^2 N_c Q_{V}}{8\pi^2 M_V^2 }\left[ g_v^2\left(\frac{5}{3}-2\epsilon\right) +g^2_a\left(\frac{5}{3}+2\epsilon\right) \right]
\end{eqnarray}in the limit $m_{q},M_V \gg m_{\mu}$. The full result for any mass regime can be found using either our full analytic expression given in Eq.\ref{full3} or using our Mathematica code. From Figs.\ref{spin1_1}-\ref{spin1_3} we conclude that $\sim$~TeV vector leptoquarks with unit couplings are not good candidates to explain the $(g-2)_{\mu}$ anomaly. If somewhat suppressed couplings (of order of $10^{-1} -10^{-2}$) are used, TeV pure vector Leptoquarks can accommodate the anomaly. In the setup where purely-chiral couplings or purely-axial couplings are presumed to be of order one the contributions are negative. Besides the $g-2_{\mu}$ anomaly bound discussed here, constraints coming from pair production searches at D0 require $M_V \gtrsim 220$~GeV \cite{vecleptoquark}.

\subsection*{Two-Particle Contributions}

\subsubsection*{Neutron Lepton and Charged Higgs}
We first present the master integral when we have a charged scalar and a new neutral fermion. First, we recognize that this scenario is equivalent to the charged scalar setup studied discussed in Section \ref{singlychargedS}, with the only difference being that the neutrino mass must be replaced by by the new neutral fermion's mass. While we continue to use the notation of a charged higgs and neutral lepton, it is also important to note that, from the point of view of this process, this scenario is exactly equivalent to a neutralino and slepton of the MSSM.

\begin{figure}[!h]
\centering
\includegraphics[scale=0.8]{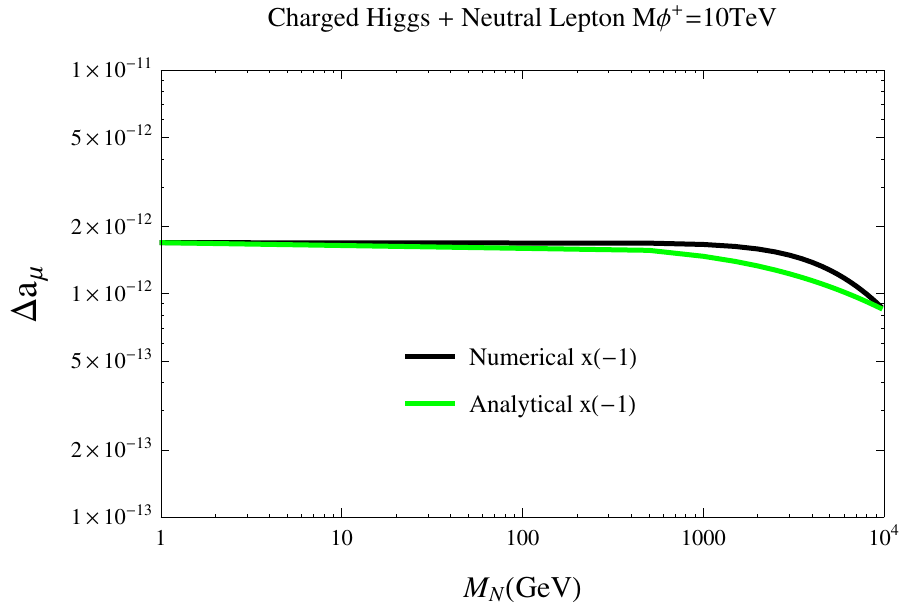}
\caption{Result for the charged higgs plus Neutral Lepton as a function of the heavy lepton mass with $M_{\phi^+}=10$~TeV.}
\label{figchargedHheavyLep}
\end{figure}

Recalling the result for a charged higgs, and making the appropriate mass replacement, we find that the contribution to the muon magnetic moment is
\begin{eqnarray}
&&
\Delta a_{\mu} (N,\phi^+) = \frac{1}{8\pi^2}\frac{m_\mu^2}{ M_{\phi^+}^2 } \int_0^1 dx \frac{g_{s2}^2 \ P_{s2}(x) + g_{p2}^2 \ P_{p2} (x) }{\epsilon^2 \lambda^2 (1-x)(1-\epsilon^{-2} x) + x},
\label{scalarmuon4}
\end{eqnarray}where 
\begin{eqnarray}
P_{s2}(x) & = &  -x (1-x)(x+\epsilon) \nonumber\\
P_{p2}(x) & = & -x (1-x)(x-\epsilon),
\label{scalarmuon5}
\end{eqnarray}with $\epsilon = M_{N}/m_{\mu}$ and $\lambda= m_{\mu}/M_{\phi^+}$. This contribution reduces to

\begin{eqnarray}
&&
\Delta a_{\mu} (N,\phi^+) = \frac{1}{4\pi^2}\frac{m_\mu^2}{ M_{\phi^+}^2 }\left\lbrace  g_{s2}^2\left(-\frac{ M_{N} }{4 m_{\mu}} -\frac{1}{12}\right) + g_{p2}^2 \left( \frac{ M_{N} }{4m_{\mu}} -\frac{1}{12}\right) \right\rbrace ,
\label{scalarmuon66}
\end{eqnarray} assuming $M_{\phi}^+ \gg m_{\mu},M_N$. Our result agrees with Ref.\cite{g2generic4}. In the regime $M_N,M_{\phi}^+ \gg m_{\mu}$ we find,

\begin{eqnarray}
&&
\Delta a_{\mu} (N,\phi^+) = \frac{g_s^2}{4\pi^2}\frac{m_\mu^2}{ M_{\phi^+}^2 }\left[ \frac{1}{6(\alpha-1)^4}\left(-2\alpha ^3-3\alpha^2 +6\alpha -1 + 6\alpha^2 Log[\alpha] \right)\right] ,
\end{eqnarray}with $g_{p2}=\pm g_{s2}$ and $\alpha=M_E^2/M_{\phi^+}^2$, which disagrees by a factor of four with the result of Ref.\cite{g-2LHC}. We show in Fig.\ref{figchargedHheavyLep} that our result agrees well the the numerical solution.

The general result can be found either numerically or applying the exact expression shown in Eq.\ref{full3} by using Mathematica code provided. We note that processes of exactly this type have been considered previously in the context of SUSY, and in at least one place in previous literature \cite{g2generic4} have been misrepresented as being dependent on the chirality of the process, when in fact left- and right-handed interactions give identical results, as has been previously known \cite{g2models4,g2generic6}.
\subsubsection*{Heavy Higgs and Charged Lepton}

\begin{figure}[!h]
\centering
\includegraphics[scale=0.8]{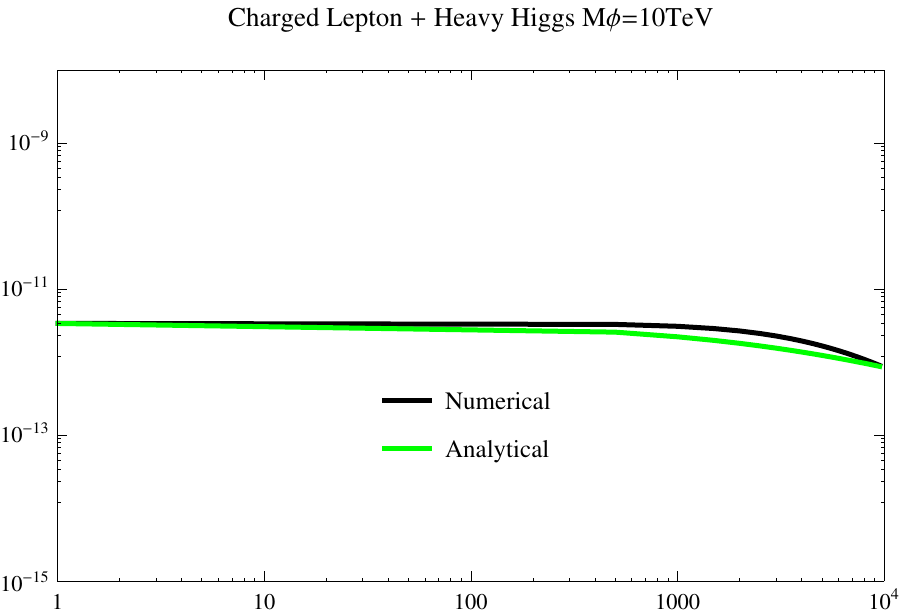}
\caption{Result for the combined contribution of the charged lepton and a heavy higgs as a function of the charged lepton mass with $M_{\Phi}=10$~TeV.}
\label{figchargedlepheavyhiggs}
\end{figure}

One more example is the scenario where we have a heavy neutral scalar and the singly charged lepton (E), which is also equivalent to the MSSM case of a sneutrino and chargino. As before we can use the master integral for the neutral scalar case discussed in Section \ref{neutralscalar} by correcting $\epsilon=1$ for that case to $\epsilon=M_E/m_{\mu}$ instead as follows:
\begin{eqnarray}
&&
\Delta a_{\mu} (\phi,E) = \frac{1}{4\pi^2}\frac{m_\mu^2}{ M_{\phi}^2 } \int_0^1 dx \frac{g_{s}^2 \ P_{s}(x) + g_{p}^2 \ P_{p}(x) }{(1-x)(1-\lambda^2 x) + \epsilon^2 \lambda^2 x}
\label{2pscalarmuon1}
\end{eqnarray}where $\lambda=m_{\mu}/M_{\phi}$ and
\begin{eqnarray}
P_{s}(x) & = &  x^2 (1-\epsilon -x)\nonumber\\
P_{p}(x) & = &  x^2 (1+\epsilon -x).
\label{2pscalarmuon2}
\end{eqnarray}
In the limit $M_E,M_{\phi} \gg m_{\mu}$ we find,
\begin{eqnarray}
&&
\Delta a_{\mu} (\phi,E) = \frac{g_s^2}{4\pi^2}\frac{m_\mu^2}{ M_{\phi}^2 }\left[ \frac{1}{6(\alpha-1)^4}\left(\alpha ^3-6\alpha^2 +3\alpha +2 + 6 \alpha Log[\alpha] \right)\right],
\label{2pscalarmuon3}
\end{eqnarray}with $g_p=\pm g_s$ and $\alpha=M_E^2/M_{\phi}^2$, which again is off by a factor of four compared to Ref.\cite{g-2LHC}. In Fig.\ref{figchargedlepheavyhiggs} one can clearly see that our result offers a good agreement the numerical one.

More complex setups (corresponding to more complete models of new physics) beyond the ones discussed here can be easily explored by making simple changes in the definition of the parameters $\epsilon$ and $\lambda$ in the integrals that we have discussed and adding together the contribution of all independent graphs. In most cases the analytic formulae provided in this section are sufficient to give the correct result, but for cases where the limits on masses considered here do not apply there are exact analytic results in the Appendix, and a Mathematica notebook is available as supplemental material to this publication (arXiv:1403.2309) which calculates the appropriate contributions numerically.

\section{Conclusions}
\label{sec:conc}

We have considered all possible models which can affect the anomalous magnetic moment of the muon at one loop with one new particle, and considered examples of cases involving multiple new particles. Analytic results have been presented for each case in relevant regions of parameter space, with exact analytic results available in the appendix. This work provides a natural reference point for understanding the implications of any new physics model for the muon magnetic moment, and corrects some misunderstandings or confusions in the previous literature.

We also provide, as supplementary material to this article, a Mathematica notebook which can be used to find the contribution of any new physics model to the muon magnetic moment. Explicit calculations are present for the models considered within this paper, and any more complete model should require only minor changes to reflect the different particle masses and charges involved.

\section*{Acknowledgements}

FQ is partly supported by the Brazilian National Counsel for Technological and Scientiﬁc Development (CNPq) and U.S.\ Department of Energy Award SC0010107. WS is partly supported by the U.S.\ Department of Energy Award SC0010107. The authors are indebted to Patrick Draper for great discussions and Prof. Michael Spira for shedding light on the calculations regarding vector leptoquarks.

\section*{Note Added}

While this article was in final preparation we noted \cite{g-2LHC}, which works in a similar vein, with a strong focus on LHC sensitivity to the various models which could explain the discrepancy in $a_\mu$.

\section{Appendix: Exact Analytical Contributions to $a_\mu$}
\label{appendix}

We present in this appendix the analytic results for each one-loop Feynman parameter integral which contributes to the anomalous magnetic moment of the muon. The integrals in question are defined as,
\begin{eqnarray}
&&
\Delta a_{\mu} ({\rm Neutral\ Scalar}) = \frac{Q_f}{8\pi^2}\frac{m_\mu^2}{ M_{\Phi}^2 } \int_0^1 dx \frac{g_{s}^2 \ P_{s}(x) + g_{p}^2 \ P_{p}(x) }{(1-x)(1-\lambda^2 x) +\epsilon^2 \lambda^2 x},
\label{generalscalar1}
\end{eqnarray}according to Fig.\ref{fig:scalar2}, with $\epsilon=m_F/m_{\mu}$, where $m_F$ being the mass of the fermion running in the loop, $\lambda=m_{\mu}/M_{\Phi}$ where $M_{\Phi}$ is the mass of the boson and,
\begin{eqnarray}
P_{s}(x) & = &  x^2 (1 + \epsilon - x),\nonumber\\
P_{p}(x) & = &  x^2 (1 - \epsilon - x).
\label{generalscalar2}
\end{eqnarray}
\begin{eqnarray}
&&
\Delta a_{\mu} ({\rm Charged\ Scalar}) = \frac{Q_S}{8\pi^2}\frac{m_\mu^2}{ M_{\Phi^+}^2 } \int_0^1 dx \frac{g_{s}^2 \ P_{s}(x) + g_{p}^2 \ P_{p} (x) }{\epsilon^2 \lambda^2 (1-x)(1-\epsilon^{-2} x) + x},
\label{generalchargedscalar1}
\end{eqnarray}according to Fig.\ref{fig:scalar1} where 
\begin{eqnarray}
P_{s}(x) & = &  -x (1-x)(x+\epsilon) \nonumber\\
P_{p}(x) & = & -x (1-x)(x-\epsilon),
\label{generalchargedscalar2}
\end{eqnarray}

\begin{eqnarray}
&&
\Delta a_{\mu} ({\rm Neutral\ Vector}) = \frac{Q_f}{8\pi^2}\frac{m_\mu^2}{ M_{\Phi}^2 } \int_0^1 dx \frac{g_{v}^2 \ P_{v}(x) + g_{a}^2 \ P_{a}(x) }{(1-x)(1-\lambda^2 x) +\epsilon^2\lambda^2 x},
\label{generalVector1}
\end{eqnarray}according to Fig.\ref{fig:vector2} where,
\begin{eqnarray}
P_{v}(x) & = &  2x(1-x)(x-2(1-\epsilon))+\lambda^2(1-\epsilon)^2 x^2 (1+\epsilon-x),\nonumber\\
P_{a}(x) & = &  2x(1-x)(x-2(1+\epsilon))+\lambda^2(1+\epsilon)^2 x^2 (1-\epsilon-x).
\label{generalVector2}
\end{eqnarray}

\begin{eqnarray}
&&
\Delta a_{\mu} ({\rm Charged\ Vector}) = \frac{Q_V}{8\pi^2}\frac{m_\mu^2}{ M_{\Phi^+}^2 } \int_0^1 dx \frac{g_{v}^2 \ P_{v}(x) + g_{a}^2 \ P_{a}(x) }{\epsilon^2\lambda^2(1-x)(1-\epsilon^{-2}x)+x},
\label{generalChargedVector1}
\end{eqnarray}according to Fig.\ref{fig:vector1} where,
\begin{eqnarray}
P_{v}(x) & = &  2x^2(1+x-2\epsilon)+\lambda^2(1-\epsilon)^2 x (1-x)(x+\epsilon),\nonumber\\
P_{a}(x) & = &  2x^2(1+x+2\epsilon)+\lambda^2(1+\epsilon)^2 x (1-x)(x-\epsilon).
\label{generalChargedVector2}
\end{eqnarray}

The above Eqs.(\ref{generalscalar1}),(\ref{generalchargedscalar1}),(\ref{generalVector1}),(\ref{generalChargedVector1}), refer to the ``neutral scalar'', ``charged scalar'', ``neutral vector'', and ``charged vector'' graphs, respectively. Note of course that it is necessary to combine the ``neutral'' and ``charged'' graphs whenever neither particle in the loop is electrically neutral. We refer to the scalar or vector running in the loop as $\Phi$ throughout this appendix, and utilize the definitions of $\lambda=\frac{m_\mu}{M_\Phi},\epsilon=\frac{m_F}{m_\mu}$, consistent with those used throughout the paper, where $F$ is the fermion running in the loop.

The scalar coupling analytic result for ``neutral scalar'' graph with a fermion of charge -1 is

\begin{align}
\Delta a_\mu=&
\frac{g_S^2m_\mu^2}{16\pi^2m_\phi^2\lambda^6}\big[-2 \lambda ^2+\lambda ^4 (2 \epsilon  (\epsilon +1)-1)+\left(\lambda ^2 \epsilon  \left(\lambda ^2 (-(\epsilon -1)) (\epsilon +1)^2+2
   \epsilon +1\right)-1\right) \log \left(\lambda ^2 \epsilon ^2\right)\nonumber\\
&+ \left(\lambda ^2
   \left(\epsilon  \left(\lambda ^4 (\epsilon -1)^2 (\epsilon +1)^3-\lambda ^2 \epsilon  (\epsilon
   +1) (3 \epsilon -1)+3 \epsilon +1\right)+1\right)-1\right) F_{N}\left(\lambda^2,\epsilon^2\right)\big]
\label{full1}   
\end{align}

where

\begin{equation}
F_{N}\left(\lambda^2,\epsilon^2\right)= \frac{2}{\sqrt{4 \lambda ^2-\left(\lambda ^2 \left(\epsilon
   ^2-1\right)-1\right)^2}}\cot ^{-1}\left(\frac{\lambda ^2
   \left(\epsilon ^2-1\right)+1}{\sqrt{4 \lambda ^2-\left(\lambda ^2 \left(\epsilon
   ^2-1\right)-1\right)^2}}\right).
\end{equation}

The pseudoscalar coupling result is given by transforming the scalar coupling result by $\epsilon\to-\epsilon$. Note that this has no effect on $F_{N}$. Naturally, either result must be scaled by the ratio of the charge of the fermion in question to the charge of -1 for a muon.

The vector coupling result for the ``neutral vector'' graph, again with a fermion of charge -1, is

\begin{align}
\Delta a_\mu & = 
\frac{g_V^2m_\mu^2}{16\pi^2m_\phi^2\lambda^6}\biggl[-4 \lambda ^2+\lambda ^4 \left(2 \epsilon ^2+2 \epsilon \right) + \lambda ^6 \left(2 \epsilon ^4-6
   \epsilon ^3+5 \epsilon ^2-1\right) \nonumber\\
   &  +\left[\lambda ^2 \left(3 \epsilon ^2+\epsilon +1\right)+\lambda ^4 \left(-4 \epsilon ^3+4 \epsilon ^2+2\epsilon -2\right) \right] \log \left(\lambda^2\epsilon^2\right)  \nonumber\\
   & +\left[ \lambda ^6 \left(-\epsilon ^6+3 \epsilon ^5-2 \epsilon ^4-2 \epsilon ^3+3 \epsilon ^2-\epsilon \right)-2\right]\log \left(\lambda^2\epsilon^2\right)\nonumber\\
   & +\left[\lambda ^2 \left(5 \epsilon ^2+\epsilon +3\right)+\lambda ^4 \left(-3 \epsilon ^4-5 \epsilon ^3+4\epsilon ^2+\epsilon -3\right) -2 \right]F_{N}\left(\lambda^2,\epsilon^2\right) \nonumber\\
   &+\left[ \lambda ^6 \left(-\epsilon ^6+7 \epsilon ^5-8 \epsilon ^4-2\epsilon ^3+3 \epsilon ^2+3 \epsilon -2\right)\right] F_{N}\left(\lambda^2,\epsilon^2\right)\nonumber\\
   &+ \left[\lambda ^8 \left(\epsilon ^8-3 \epsilon
   ^7+\epsilon ^6+5 \epsilon ^5-5 \epsilon ^4-\epsilon ^3+3 \epsilon ^2-\epsilon \right)\right]F_{N}\left(\lambda^2,\epsilon^2\right)\biggr], \nonumber\\
\label{full2}
\end{align}and the axial vector result is once more obtained by taking $\epsilon\to-\epsilon$.

%\begin{align}
%\Delta a_\mu=&
%\frac{g_V^2m_\mu^2}{16\pi^2m_\phi^2\lambda^6}\big[-4 \lambda ^2+\lambda ^4 \left(2 \epsilon ^2+2 \epsilon \right)+\lambda ^6 \left(2 \epsilon ^4-6
%   \epsilon ^3+5 \epsilon ^2-1\right)
%   +\left(\lambda ^2 \left(3 \epsilon ^2+\epsilon +1\right)+\lambda ^4 \left(-4 \epsilon ^3+4 \epsilon ^2+2    \epsilon -2\right)+ \lambda ^6 \left(-\epsilon ^6+3 \epsilon ^5-2 \epsilon ^4-2 \epsilon ^3+3 \epsilon ^2-\epsilon \right)-2\right)\log \left(\lambda^2\epsilon^2\right)\nonumber\\
%&+\left(\lambda ^2 \left(5 \epsilon ^2+\epsilon +3\right)+\lambda ^4 \left(-3 \epsilon ^4-5 \epsilon ^3+4
%   \epsilon ^2+\epsilon -3\right)+\lambda ^6 \left(-\epsilon ^6+7 \epsilon ^5-8 \epsilon ^4-2
%   \epsilon ^3+3 \epsilon ^2+3 \epsilon -2\right)+\lambda ^8 \left(\epsilon ^8-3 \epsilon
%   ^7+\epsilon ^6+5 \epsilon ^5-5 \epsilon ^4-\epsilon ^3+3 \epsilon ^2-\epsilon \right)-2\right)F_{N}\left(\lambda^2,\epsilon^2\right)
%\big]
%\label{full2}
%\end{align}

The ``charged scalar'' graph, for a scalar with unit charge, gives

\begin{align}
\Delta a_\mu=&
\frac{g_S^2m_\mu^2}{16\pi^2m_\phi^2\lambda^6}\big[\lambda ^4 \left(2 \epsilon ^2+2 \epsilon +1\right)-2 \lambda ^2+\left(\lambda ^2 \left(2 \epsilon ^2+\epsilon +1\right)+\lambda ^4 \left(-\epsilon ^4-\epsilon
   ^3\right)-1\right) \log \left(\lambda ^2 \epsilon^2\right)\nonumber\\
&\left(\lambda ^2 \left(3 \epsilon ^2+\epsilon +2\right)+\lambda ^4 \left(-3 \epsilon ^4-2 \epsilon
   ^3-\epsilon ^2-\epsilon -1\right)+\lambda ^6 \left(\epsilon ^6+\epsilon ^5-\epsilon ^4-\epsilon
   ^3\right)-1\right)F_{C}\left(\lambda^2,\epsilon^2\right)
\big],
\label{full3}
\end{align}

where

\begin{equation}
F_{C}\left(\lambda^2,\epsilon^2\right)= \frac{2}{\sqrt{\lambda ^4 \left(-\left(\epsilon ^2-1\right)^2\right)+2 \lambda ^2
   \left(\epsilon ^2+1\right)-1}} \cot ^{-1}\left(\frac{\lambda ^2 \left(\epsilon ^2-1\right)+1}{\sqrt{\lambda ^4
   \left(-\left(\epsilon ^2-1\right)^2\right)+2 \lambda ^2 \left(\epsilon
   ^2+1\right)-1}}\right).
\end{equation}

As usual, the pseudoscalar case is identical to the scalar case with $\epsilon\to-\epsilon$, and $F_{C}$ is invariant under this change. This result will need to be scaled by the ratio of the scalar's charge to 1.

The final graph is the ``charged vector'' case, which gives

\begin{align}
\Delta a_\mu=&
\frac{g_V^2m_\mu^2}{16\pi^2m_\phi^2\lambda^6}\biggl[
\left[-2 + \left(7 - 6 \epsilon + 5 \epsilon^2\right) \lambda^2 + \left(-5 + 5 \epsilon - 
    5 \epsilon^2 + 7 \epsilon^3 - 4 \epsilon^4\right) \lambda^4 \right]\log \left(\lambda ^2 \epsilon^2\right)\nonumber\\
     &+ \left[ \left(\epsilon - 2 \epsilon^2 + \epsilon^3\right) \lambda^5 + \left(\epsilon - 2 \epsilon^2 +  2 \epsilon^3 - \epsilon^4 - \epsilon^5 + \epsilon^6\right) \lambda^6 \right] \log \left(\lambda ^2 \epsilon
   ^2\right)\nonumber\\
&-\left[2 + \left(-9 + 6 \epsilon - 7 \epsilon^2\right) \lambda^2 
+ \left(12 - 11 \epsilon +  13 \epsilon^2 - 13 \epsilon^3 + 9 \epsilon^4\right)\lambda^4 \right]F_C\left(\lambda^2,\epsilon^2\right)\nonumber\\
&-\left[ \left(-\epsilon + 2 \epsilon^2 - \epsilon^3\right) \lambda^5 
+ \left(-5 + 4 \epsilon + 2 \epsilon^2 - 2 \epsilon^3 - 2 \epsilon^4 
+ 8 \epsilon^5 - 5 \epsilon^6\right) \lambda^6 \right]F_C\left(\lambda^2,\epsilon^2\right)\nonumber\\
&-\left[ \left(\epsilon - 2 \epsilon^2 + 2 \epsilon^3 -2 \epsilon^4 + \epsilon^5\right) \lambda^7 + \left(\epsilon - 2 \epsilon^2 + \epsilon^3 - \epsilon^4 + 3 \epsilon^5 - 2 \epsilon^6 - \epsilon^7 + \epsilon^8\right) \lambda^8\right] F_C\left(\lambda^2,\epsilon^2\right)\nonumber\\
&-4 \lambda^2 + \left(12 - 12 \epsilon + 6 \epsilon^2\right) \lambda^4 + \left(-1 + 
    \epsilon^2 + 2 \epsilon^3 - 2 \epsilon^4\right) \lambda^6\biggr]
\label{full4}    
\end{align}

%\begin{align}
%\Delta a_\mu=&
%\frac{g_V^2m_\mu^2}{16\pi^2m_\phi^2\lambda^6}\biggl[
%\left[-2 + \left(7 - 6 \epsilon + 5 \epsilon^2\right) \lambda^2 + \left(-5 + 5 \epsilon - 
%    5 \epsilon^2 + 7 \epsilon^3 - 4 \epsilon^4\right) \lambda^4 + \left(\epsilon - 
%    2 \epsilon^2 + \epsilon^3\right) \lambda^5 + \left(\epsilon - 2 \epsilon^2 + 
%    2 \epsilon^3 - \epsilon^4 - \epsilon^5 + \epsilon^6\right) \lambda^6 \right] \log \left(\lambda ^2 \epsilon
%   ^2\right)\nonumber\\
%&-\left(2 + \left(-9 + 6 \epsilon - 7 \epsilon^2\right) \lambda^2 + \left(12 - 11 \epsilon + 
%    13 \epsilon^2 - 13 \epsilon^3 + 9 \epsilon^4\right) \lambda^4 + \left(-\epsilon + 
%    2 \epsilon^2 - \epsilon^3\right) \lambda^5 + \left(-5 + 4 \epsilon + 
%    2 \epsilon^2 - 2 \epsilon^3 - 2 \epsilon^4 + 8 \epsilon^5 - 
%    5 \epsilon^6\right) \lambda^6 + \left(\epsilon - 2 \epsilon^2 + 2 \epsilon^3 - 
%    2 \epsilon^4 + \epsilon^5\right) \lambda^7 + \left(\epsilon - 2 \epsilon^2 + 
%    \epsilon^3 - \epsilon^4 + 3 \epsilon^5 - 2 \epsilon^6 - \epsilon^7 + 
%    \epsilon^8\right) \lambda^8\right) F_C\left(\lambda^2,\epsilon^2\right)\nonumber\\
%&-4 \lambda^2 + \left(12 - 12 \epsilon + 6 \epsilon^2\right) \lambda^4 + \left(-1 + 
%    \epsilon^2 + 2 \epsilon^3 - 2 \epsilon^4\right) \lambda^6\biggr]
%\label{full4}    
%\end{align}

Once again, the axial vector case is equivalent with $\epsilon\to-\epsilon$. All of these results are available for use in a Mathematica notebook file which is online as supplementary material to this paper available on arXiv:1403.2309. Note that these analytic expressions have very large cancellations in terms, and therefore high numerical precision is needed to achieve the correct result.

\end{document}